\documentclass[twocolumn]{aastex62}
\usepackage{amsmath}
\usepackage{colortbl}

\usepackage{etoolbox}
\makeatletter 
  \patchcmd{\NAT@citex}
    {\@citea\NAT@hyper@{%
      \NAT@nmfmt{\NAT@nm}%
      \hyper@natlinkbreak{\NAT@aysep\NAT@spacechar}{\@citeb\@extra@b@citeb}%
      \NAT@date}}
    {\@citea\NAT@nmfmt{\NAT@nm}%
    \NAT@aysep\NAT@spacechar\NAT@hyper@{\NAT@date}}{}{}

  \patchcmd{\NAT@citex}
    {\@citea\NAT@hyper@{%
      \NAT@nmfmt{\NAT@nm}%
      \hyper@natlinkbreak{\NAT@spacechar\NAT@@open\if*#1*\else#1\NAT@spacechar\fi}%
        {\@citeb\@extra@b@citeb}%
      \NAT@date}}
    {\@citea\NAT@nmfmt{\NAT@nm}%
    \NAT@spacechar\NAT@@open\if*#1*\else#1\NAT@spacechar\fi\NAT@hyper@{\NAT@date}}
    {}{}
\makeatother

\newcommand{\flash}{\texttt{FLASH}}
\newcommand\mesa{\texttt{MESA}}

\usepackage{color,soul}    %
\soulregister\citet7
\soulregister\citep7
\soulregister\ref7
\newtoggle{referee}            %
\togglefalse{referee}          %

\iftoggle{referee}{
   \usepackage{color,soul}    %
   \soulregister\citet7
   \soulregister\citep7
   \soulregister\ref7
   \newcommand{\hlm}[1]{%
       \hl{#1}
   }                          %
}{
   \newcommand{\hlm}[1]{{#1}} %
}                              %

\graphicspath{{./}{figures/}}

\received{May 30, 2022}
\revised{July 19, 2022}
\accepted{July 21, 2022}
\submitjournal{ApJ}

\shorttitle{RSG Pre-SN Outbursts in 3D}
\shortauthors{Tsang, Kasen, and Bildsten}

\begin{document}

\title{3D Hydrodynamics of Pre-supernova Outbursts in Convective Red Supergiant Envelopes}

\correspondingauthor{Benny T.-H. Tsang}
\email{benny.tsang@berkeley.edu}

\author{Benny T.-H. Tsang}
\affiliation{Department of Astronomy and Theoretical Astrophysics Center, University of California, Berkeley, CA 94720, USA}
\affiliation{Kavli Institute for Theoretical Physics, University of California, Santa Barbara, CA 93106, USA}

\author{Daniel Kasen}
\affiliation{Department of Astronomy and Theoretical Astrophysics Center, University of California, Berkeley, CA 94720, USA}
\affiliation{Department of Physics, University of California, Berkeley, CA 94720, USA}
\affiliation{Lawrence Berkeley National Laboratory, Berkeley, CA 94720, USA}

\author{Lars Bildsten}
\affiliation{Kavli Institute for Theoretical Physics, University of California, Santa Barbara, CA 93106, USA}
\affiliation{Department of Physics, University of California, Santa Barbara, CA 93106, USA}
\nocollaboration



\begin{abstract}
Eruptive mass loss likely produces the energetic outbursts observed from some massive stars before they undergo core-collapse supernovae (CCSNe). 
The resulting dense circumstellar medium (CSM) may also cause the subsequent SNe to be observed as Type IIn events. 
The leading hypothesis of the cause of these outbursts is the response of the envelope of the 
red supergiant (RSG) progenitor to energy deposition 
in the months to years prior to collapse. 
Early theoretical studies of this phenomena were limited to 1D, leaving the 3D convective RSG structure unaddressed. 
Using \texttt{FLASH}'s hydrodynamic capabilities, we explore the 3D outcomes by constructing convective RSG envelope models and depositing energies less than the envelope binding energies on timescales shorter than the envelope dynamical time deep within them. 
We confirm the 1D prediction of an outward moving acoustic pulse steepening into a shock, unbinding the outermost parts of the envelope.
However, we find that the initial $2-4$\,km\,s$^{-1}$ convective motions seed the intrinsic convective instability associated with the high entropy material deep in the envelope, enabling gas from deep within the envelope to escape, increasing the amount of ejected mass compared to an initially `quiescent' envelope.
The 3D models reveal a rich density structure, with column densities varying by $\approx$10$\times$ along different lines of sight. 
Our work highlights that the 3D convective nature of RSG envelopes impacts our 
ability to reliably predict the outburst dynamics, the amount, and the spatial
distribution of the ejected mass associated with deep energy deposition.

\end{abstract}


\section{Introduction} 
\label{sec:intro}

Interaction-powered SNe are a diverse collection of transient objects characterized by signs of interaction between the exploding star and a dense CSM. 
Three fundamental questions concern this diverse group of objects --  
how has the CSM come into place? How is the CSM related to the properties of the progenitor star system? And how does its presence impact the subsequent SN emission?
Ongoing and future missions will likely reveal a wide variety of mass eruptions and explosions from massive stars. A comprehensive understanding of the origins and consequences of the mass loss is crucial in reliably interpreting observations and inferring massive star properties. 

Type IIn SNe are the prototypical class of interaction-powered SN. They are characterized by narrow hydrogen emission lines, commonly interpreted as the result of CCSNe exploding into the dense and slow-moving CSM just above the dying stars.
Recent observational surveys reported that a substantial fraction of IIn SNe exhibited bright outbursts months to years before the actual SN explosions \citep{Ofek14,Strotjohann21}, suggesting the outburst-driving mechanisms could also produce the CSM that gives rise to the subsequent IIn SNe.
In their samples, the outbursts carried energies in the range of $10^{46-49}$\,erg. They contributed about $10^{-2}$ to a few $M_{\odot}$ of gas to the CSM, typically at radii of about $\sim10^{14}-10^{15}$\,cm. Most of these SN precursors are luminous at $10^{6-8}$\,$L_{\odot}$.
In recent years, flash spectroscopy also provides a direct way to probe the physical conditions of the CSM immediately outside the more common Type II-P and II-L SNe \citep{Kangas16,Yaron17, Bruch21}.

More generally, even the more common Type II-P and II-L SNe have shown signs of interactions \citep{Mauerhan13,Benetti16,Kangas16}.
For example, the long plateau and high late-time luminosity of the SN 2017gmr are suggestive of CSM interactions \citep{Andrews19}.
Initially categorized as IIn SNe by narrow hydrogen emission lines, SN 2013fs and 2013fr later resembled behaviors closer to the normal Type II-P and II-L \citep{Bullivant18}.
This blending of classes strongly suggests that outbursts and mass loss are likely common in the late stages of massive star evolution, and that substantial mass loss is not a special feature exclusive for IIn SNe.

Numerous models have examined the potential physical mechanisms responsible for the outbursts from massive stars, but the theoretical landscape is still unclear.
The four main proposed mechanisms include
dissipation of internal gravity waves driven by core burning \citep{QS12,SQ14,Fuller17,RM17,WF21,WF22},
unstable late-stage nuclear burning triggered by turbulent convection \citep{SA14},
pulsational pair-instability leading to explosive burning \citep{Woosley07},
and binary interaction depositing gravitational energy onto the stellar envelope \citep{Chevalier12}.

In order to understand pre-SN outbursts and the associated mass eruptions, many theoretical studies have explored how RSG envelopes react to energy deposition hydrodynamically \citep{Dessart10,Quataert16,Kuriyama20,Morozova20,Ko22,Linial21}.
{Based on various 1D stellar models, their main conclusions are:
(1) energy injection comparable to the RSG envelope's binding energy can lead to mass ejection and CSM formation;
(2) the energy injection timescale relative to the RSG envelope's dynamical time determines the mode of mass loss, e.g., as shock-induced episodic eruptions or more steady winds;
(3) the general setup of energy injection within stellar envelopes alike can explain some of the photometric, spectral, and flash-ionization observations.}

While most theoretical studies modeled energy deposition and mass-erupting outbursts in 1D, polarimetry observations have shown that interacting SNe can be considerably asymmetric \citep{patat11,Levesque14,Mauerhan14,Reilly17}. Their diverse light curve and spectral features also point to intrinsically complex CSM geometries \citep{Kiewe12,Soumagnac19,Nyholm20,Ransome21}.
{In fact, 3D radiation hydrodynamical simulations of RSG envelopes have shown that convection can produce shocks and inhomogeneous large eddies near the surface \citep{Chiavassa09,Chiavassa11}, and that above a certain critical radius radiative loss in the convective layer can render the usual mixing length theory (MLT) description inaccurate \citep{Goldberg21}.}
Overall, 1D and simplistic stellar models may be missing important physical ingredients for our understanding of mass loss, stability, and explosions of massive stars.

\citet{Leung20} have modeled energy deposition on a RSG envelope using 2D hydrodynamical simulations. They found that Rayleigh-Taylor instabilities (RTIs) could smooth out density inversions, and model asymmetries seemed to play a minor role during envelope expansion. {However, they did not include convection in their initial model. We are still uncertain how a convective RSG envelope would react to energy deposition beyond 1D.}

In this work, we simulate the hydrodynamics of pre-SN outbursts from RSG envelopes in 3D for the first time. Our core motivation is to shed light on the gas dynamics of a convective stellar envelope when it is subjected to an explosive energy deposition. The 3D models will also reveal how the envelope material is ejected to create the CSM and what the resultant density structures look like.
We systematically vary the model parameters to study how mass ejection depends on the properties of the envelope model, the amount and rate of energy deposition, and model dimension.

In Section \ref{sec:methods}, we describe the numerical methods, detailing the new modifications required for our simulation suite.
In Section \ref{sec:results}, we present the results from our simulations. We showcase the vital differences in outburst behavior with and without including stellar convection in 3D (\ref{sec:conv_role}). We then compare models performed in 1D and 3D with otherwise identical initial conditions (\ref{sec:1D3D}). We examine the dependence on energy deposition parameters in Section \ref{sec:dep_study}. In Section \ref{sec:3dejecta}, we reveal the 3D spatial structure of the ejecta mass.
In Section \ref{sec:summary}, we summarize our results and discuss their implications. We discuss limitations of our models and suggest some future directions in Section \ref{sec:caveats}.

\section{Numerical Methods}
\label{sec:methods}

We use two sets of numerical tools to simulate the hydrodynamics of pre-SN outbursts in a RSG envelope.
The progenitor model is generated using Modules for Experiments in Stellar Astrophysics \mesa\ \citep{Paxton2011, Paxton2013,Paxton2015,Paxton18,Paxton19}. The energy deposition, shock propagation, envelope dynamics, mass loss, and the resultant ejecta properties are followed using the adaptive mesh refinement (AMR) code \flash\ \citep{Fryxell00}, version 4.6.2.
Technical details of the numerical setup will be described below. 

\begin{deluxetable}{cc}[t]
\tablecaption{
Properties of the RSG progenitor model. 
\label{tab:rsg_progenitor}}
\tablecolumns{3}
\tablewidth{10pt}
\tablehead{
\colhead{Quantity } & \colhead{Value}
}
\startdata
$M_{\rm ZAMS}$ & 15.0\,$M_{\odot}$ \\
$M_{*}$ & 14.5\,$M_{\odot}$ \\
$M_{\rm He, c}$ & 5.1\,$M_{\odot}$ \\
$M_{\rm env}$ & 6.5\,$M_{\odot}$ \\
$E_{\rm bind}$ & $-7.54 \times 10^{47}$\,{\rm erg}\\
$R_{*}$  & 721.2\,$R_{\odot}$ \\
$T_{\rm eff}$  & 3903\,K \\
$L_{*}$ & 1.04 $\times 10^{5}$\,$L_{\odot}$ \\ 
$Z$ & 0.02 \\
$t_\textrm{dyn} = \sqrt{R_{*}^{3}/G M_{*}}$ & 8.1 $\times 10^{6}$\,s \\
$v_\textrm{esc, surf}$ & 87.6\,km\,s$^{-1}$
\enddata
\tablenotetext{}{\textbf{Note:} The stellar properties are defined as follows. 
(1) Zero-age main-sequence (ZAMS) mass $M_{\rm ZAMS}$, 
(2) final progenitor mass $M_{*}$, 
(3) helium core mass $M_{\rm He, c}$, 
(4) mass in the hydrogen envelope above 200\,$R_{\odot}$,
(5) gravitational binding energy,
(6) stellar radius $R_{*}$, 
(7) surface temperature $T_{\rm eff}$,
(8) metallicity $Z$,
(9) global dynamical time $t_{\rm dyn}$,
(10) escape velocity on the envelope surface.}
\end{deluxetable}

\subsection{Progenitor Model in \mesa}

As the initial condition of the RSG outburst simulations, we choose a representative massive star model from the \mesa\ model suite of \citet{Goldberg20}. The chosen model is named `M12.7\_R719\_E0.84', a non-rotating star with a zero-age main sequence (ZAMS) mass of 15\,$M_{\odot}$. The inlist parameters for the model follow \texttt{example\_make\_pre\_ccsn} in the \mesa\  \texttt{test\_suite}. Revised values were used in the mixing length in the the hydrogen-rich envelope $\alpha_{\rm env} = 3.0$, core overshooting parameter $f_{\rm ov} = 0.01$, and wind efficiency $\eta_{\rm wind} = 0.2$. 
We ran the model until core collapse, when mass loss during stellar evolution has reduced the final stellar mass to 14.5\,$M_{\odot}$. We then perform the model hand-off to \flash.
We checked that the variations in the hydrogen-rich envelope of the RSG are minimal in the last decade of stellar evolution. 
The envelope structure we adopt therefore serves as a reasonable initial condition for modeling pre-SN outburst activities on RSGs.
Key properties of the progenitor model are summarized in Table \ref{tab:rsg_progenitor}.

\subsection{\mesa-to-\flash\ hand-off}

The hydrodynamical simulations are set up in 3D spherical geometry in \flash.
\hlm{We fix the simulation domain's inner boundary at $R_{\rm inner} = 200$\,$R_{\odot}$, inside the hydrogen-rich envelope and above the helium core. As our main goal is to expose the effects of convection on shock propagation in the outer envelope, the inner region near the helium core is excluded for computational expediency.}
The outer boundary is set at $R_{\rm outer} = 5000$\,$R_{\odot}$, far enough to capture the intrinsically aspherical mass loss.
The angular dimensions span $\theta \in [\pi/4, 3\pi/4]$ and $\phi \in [0, 2\pi]$. Omission of the polar regions helps to avoid excessively small cell sizes and time steps.
In this paper, physical quantities such as mass and energy from the 3D models are re-scaled to correspond to a full stellar envelope with $4\pi$ solid angle coverage.
The boundary conditions of the inner and outer radial directions are reflective and outflow, respectively. 
Boundaries are reflective (periodic) in the polar (azimuthal) direction\footnote{In Appendix E of \citet{Leung20}, the authors have reported that using a reflective boundary condition in the $\theta$ direction can moderately suppress $\theta$-velocities near the inner boundary, but the effects on the outer part of the stellar envelope are minimal.}.

At the highest refinement level, the AMR grid has an effective resolution of $(N_{\rm r}, N_{\theta}, N_{\phi}) = (1280, 256, 256)$. The levels of refinement are constrained to have radial cell sizes of $\Delta r/r \lesssim 4\%$.
To initialize the progenitor envelopes on \flash's 3D grid, we linearly interpolate \mesa's density and temperature profiles based on the cell's radial coordinate at $R_{\rm inner} \le R \le R_{\rm outer}$. 
Above $R_{\rm inner}$, the stellar envelope has a total mass of $M_{\rm env} = 6.5$\,$M_{\odot}$.
The core region inward of $R_{\rm inner}$ has a mass of $M_{\rm core} = 8.0\,M_{\odot}$ and is treated as a point mass with gravity (Section \ref{sec:gravity}).
With the chosen resolution, the stellar envelope is resolved by $\approx$192 cells radially in the initial \flash\ setup.
In the region outside the envelope ($R > R_{*}$), we impose a constant density floor of $10^{-14}$\,g\,cm$^{-3}$. The total gas mass introduced by imposing the density floor outside $R_{*}$ is $\lesssim 10^{-3}$\,$M_{\odot}$, which is much smaller than the typical ejecta mass of $M_{\rm ej} \sim 0.1 M_{\odot}$ following energy deposition (Section \ref{sec:conv_role}).
In the following, we refer to the ejected gas launched from the stellar envelope following energy deposition as the \emph{ejecta}, which occurs before the actual supernova explosions.

\subsection{Energy Deposition, Gravity, and Hydrodynamics}
\label{sec:gravity}
We adopt an agnostic approach to the driving mechanisms of the outbursts.
Energy deposition is implemented as an artificial, constant-rate thermal heating in a selected region of the simulation domain. The numerical implementation is based on modifying the \texttt{Heat} source term unit in \flash. This heating term is applied explicitly following the hydrodynamics update.
To maintain numerical stability in the explicit heating scheme, we adopt a Courant factor of $\mathcal{C} = 0.3$. In other words, the time step limit for energy deposition, $\Delta t_{\rm heat} = \mathcal{C}\, \textrm{min}\left[E_{\rm int, i} / L_{\rm dep, i}\right]$, is set such that the fractional change in internal energy in any cell can at most be 30\%. Here $E_{\rm int, i}$ and $L_{\rm dep, i}$ are the total internal energy and the local energy deposition rate of the $i$-th cell.
To facilitate comparison, we follow previous studies to deposit energy near the bottom of the stellar envelopes. The total energy deposition rate $L_{\rm dep} = \sum L_{\rm dep,i}$ is chosen as a fraction of the gravitational binding energy of the model envelope. 

The stellar core inside the hydrogen-rich envelope only interacts with the envelope through gravity as a point mass, which is implemented in the spherically symmetric gravity unit \texttt{PlanePar} in \flash.
Since the envelope mass is comparable to that of the core, it is crucial to account for the self-gravity of the envelope. Hydrostatic equilibrium in the envelope would otherwise not be maintained.
To this end, we modify the gravity unit to compute gravitational acceleration using $\mathbf{a}_{\rm grav}(r) = -G m(r)/r^2 \hat{r}$, where $G$ is the gravitational constant, $r$ is the radius, and $m(r)$ is the total mass of the core \emph{and} the envelope mass enclosed inside $r$.
This implementation effectively assumes a spherically symmetric mass distribution. As we will see in Section \ref{sec:conv_role}, the 3D gas distributions in the RSG envelope following energy deposition are not perfectly symmetric. We have computed the exact gravitational acceleration by integrating the mass distribution at a few characteristic times. We found that the deviations are at most $\lesssim 5-10$\%, verifying that our approximate treatment of self-gravity is acceptable.

\begin{figure}
\centering
\includegraphics[width=\columnwidth]{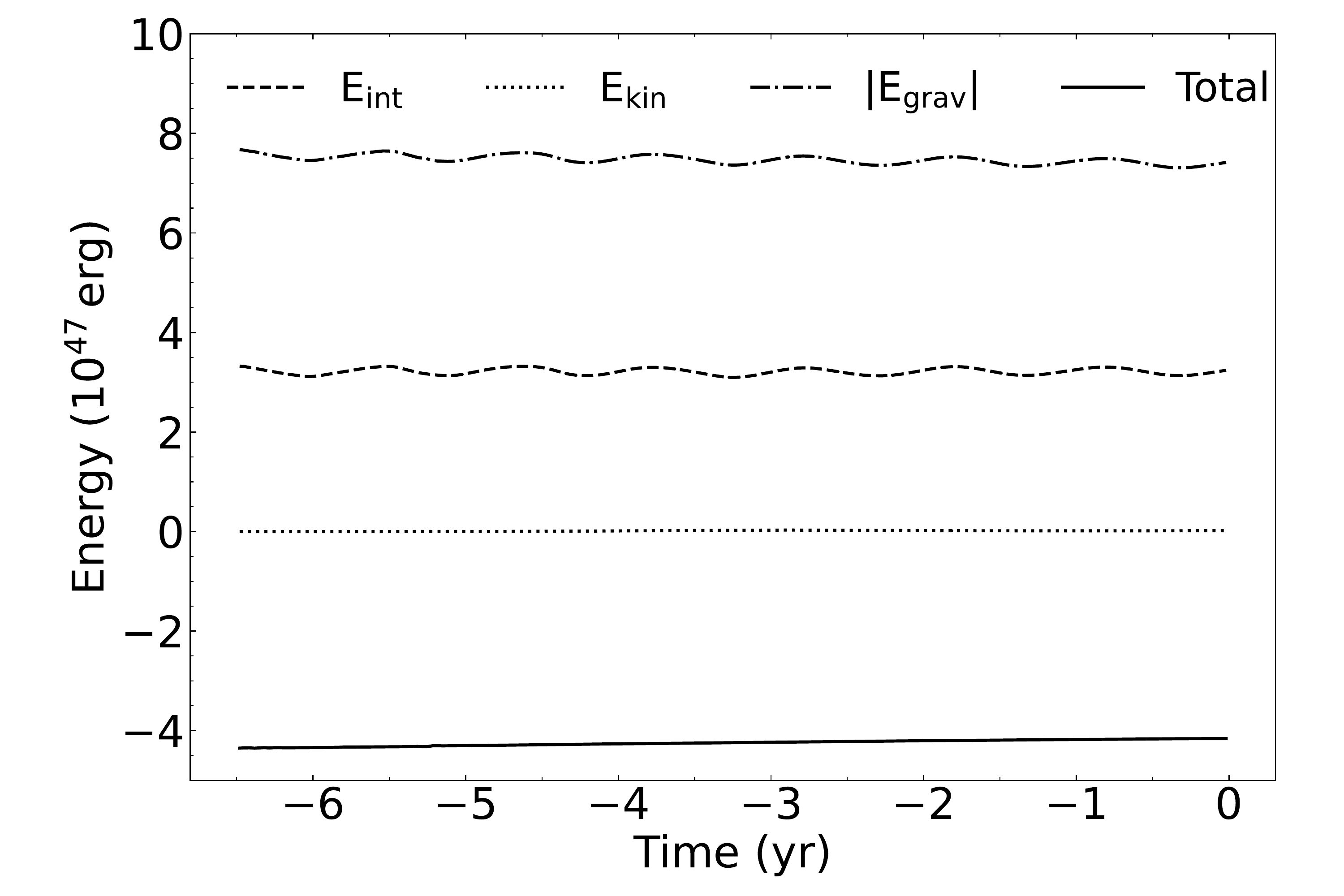}
\caption{Time evolution of different forms of energy during the preparation of the 3D convective equilibrium model.} \label{fig:E_time_prep}
\end{figure}

Hydrodynamics is followed using the \texttt{HLLC} Riemann solver. The \texttt{Helmholtz} equation of state (EoS) \citep{TS00} is used to determine the thermodynamics in the envelope, which assumes complete ionization of all species and includes pressure contributions from both ionized gas and radiation.
The original \texttt{Helmholtz} EoS table spans density and temperature ranges of $10^{-12} \leq \rho \leq 10^{15}$\,g\,cm$^{-3}$ and $3 \leq \log(T/K) \leq 13$. In order to accommodate the density floor at $10^{-14}$\,g\,cm$^{-3}$, the EoS is extended beyond the lower density limit by assuming an ideal gas law for electrons and ions \citep[following][]{Fernandez18}. 
Gas remains fully ionized in the majority of the progenitor envelope. 
A single species with the mean atomic weight of the \mesa\ hydrogen envelope is used in the \flash\ simulations, which effectively neglects the partial ionization of hydrogen near the top of the envelope ($R \gtrsim 650\,R_{\odot}$), see Section \ref{sec:caveats}.

\begin{figure*}
\centering
\includegraphics[width=\columnwidth]{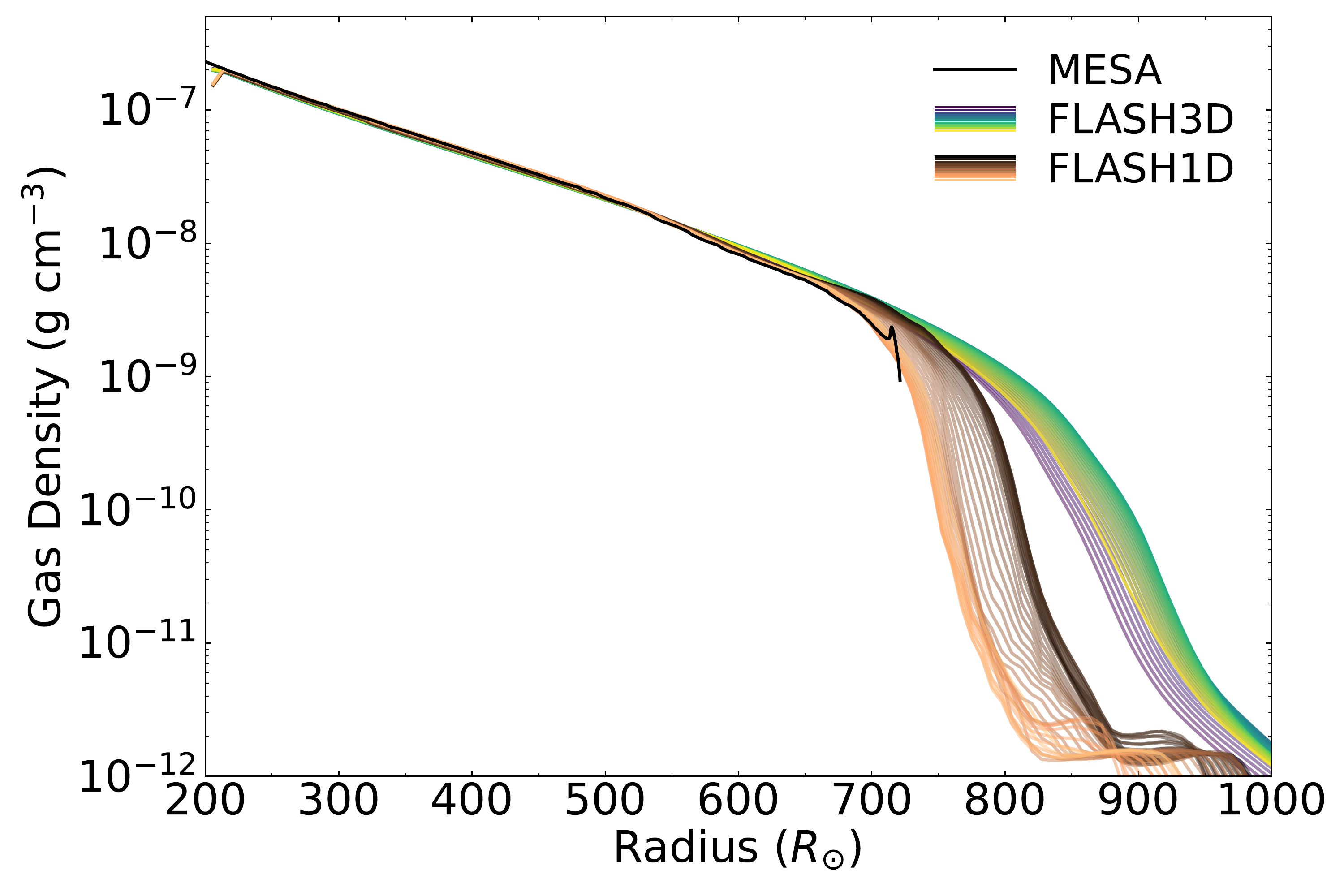}
\includegraphics[width=\columnwidth]{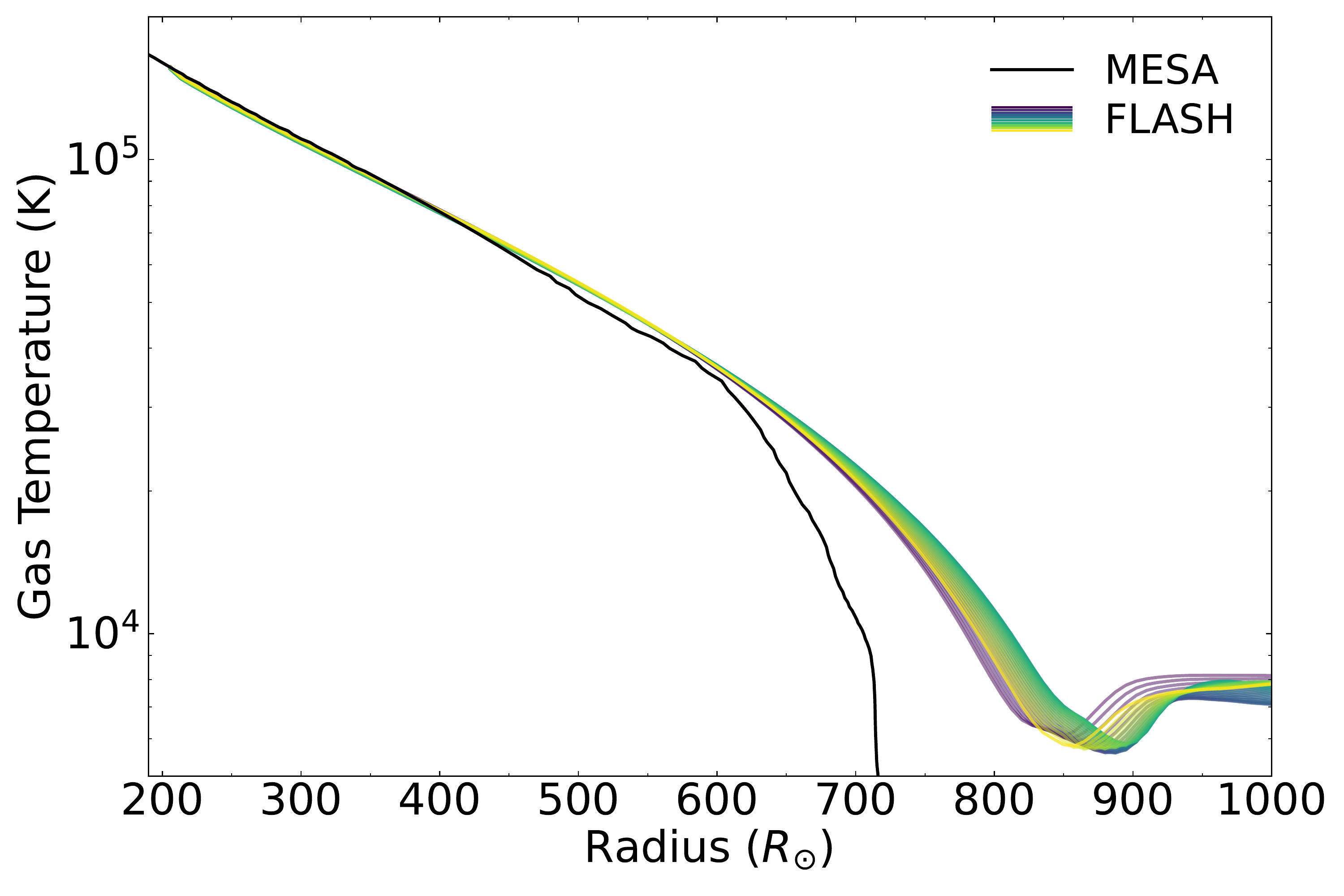}
\includegraphics[width=\columnwidth]{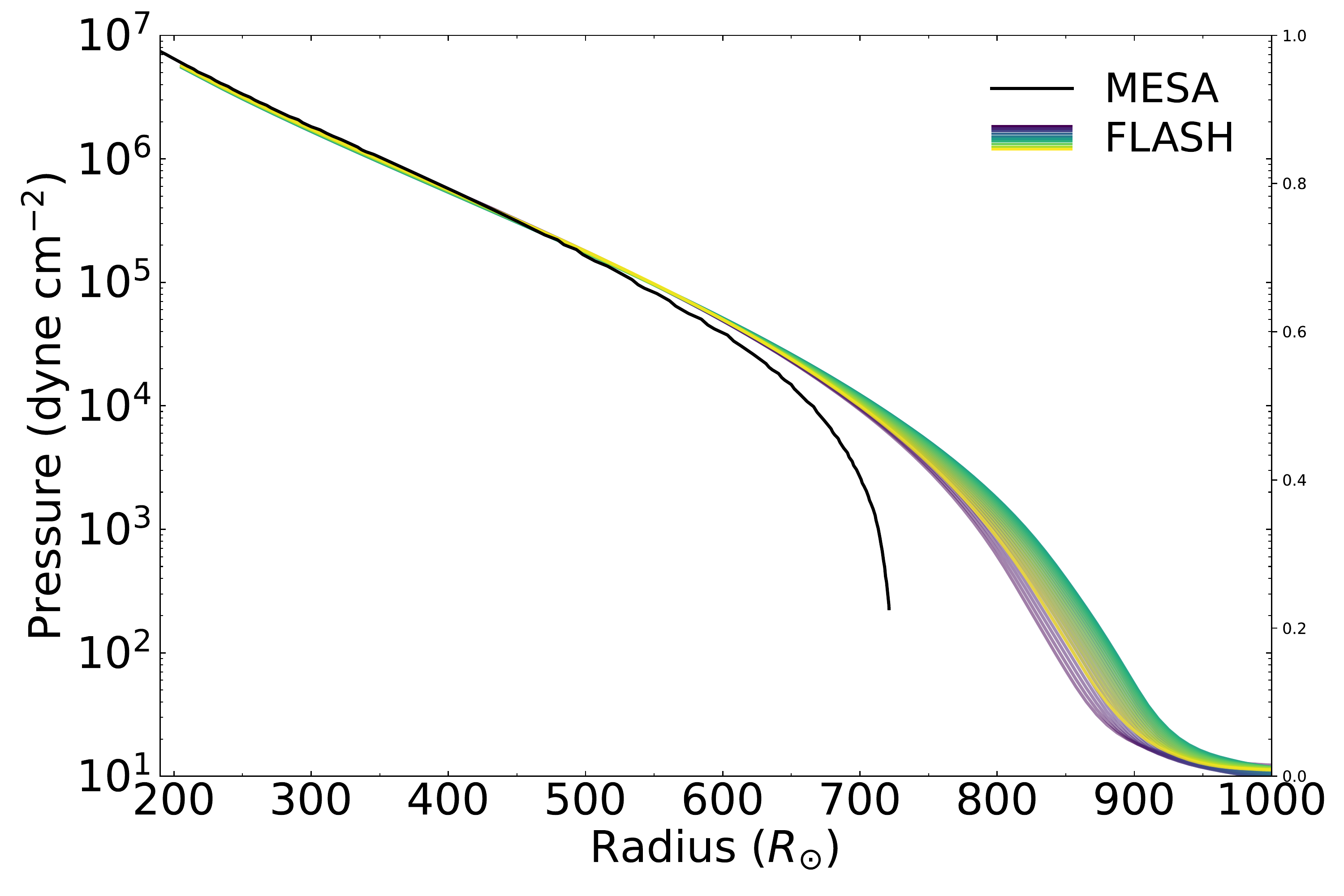}
\includegraphics[width=\columnwidth]{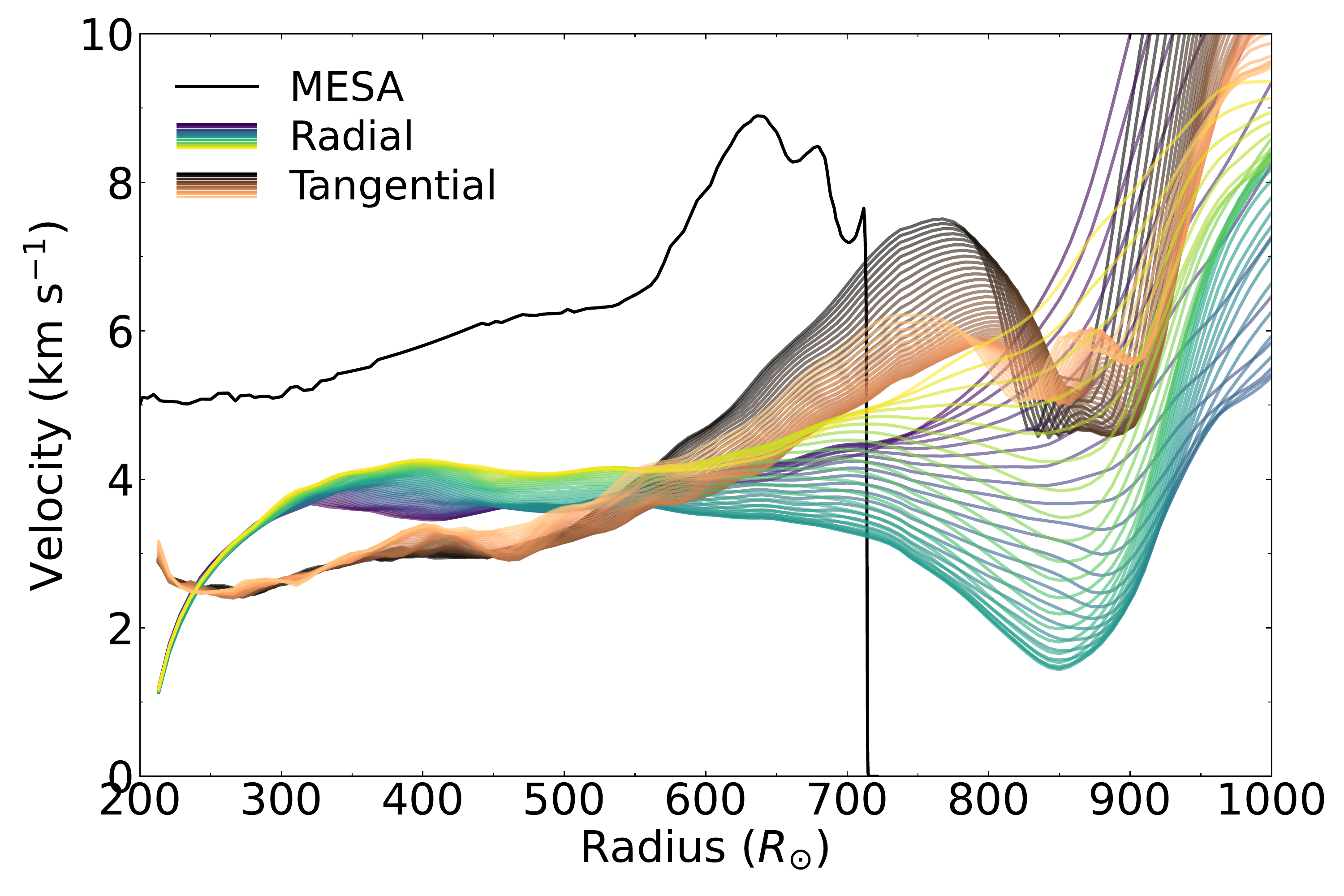}
\caption{Radial profiles of density, temperature, pressure (radiation and gas), and velocity in the 3D convective equilibrium model. The color lines show the last 5\,$t_{\rm dyn}$ of evolution at an interval of 0.1\,$t_{\rm dyn}$ before energy deposition.  
The black curves in the density, temperature, and pressure panels are the 1D \mesa\ profiles used to initialize the 3D \flash\ model.
In the velocity panel, the black curve denotes the convective velocity predicted by the MLT in \mesa.
The density profiles from the 1D control run (FLASH1D) is also shown for comparison. 
} \label{fig:conv_eqm_profs}
\end{figure*}

\begin{figure*}
\centering
\includegraphics[width=0.49\textwidth]{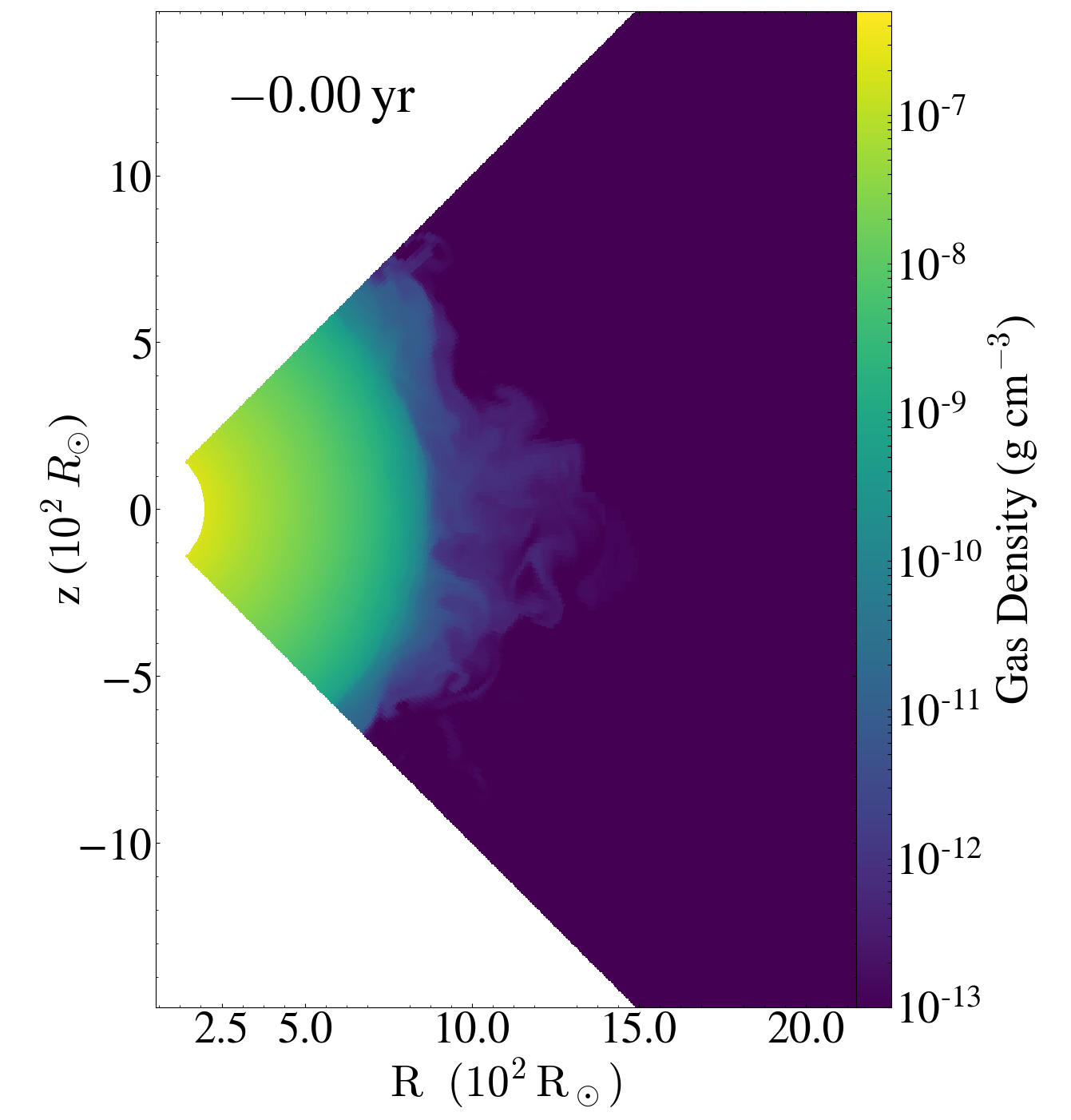}
\includegraphics[width=0.49\textwidth]{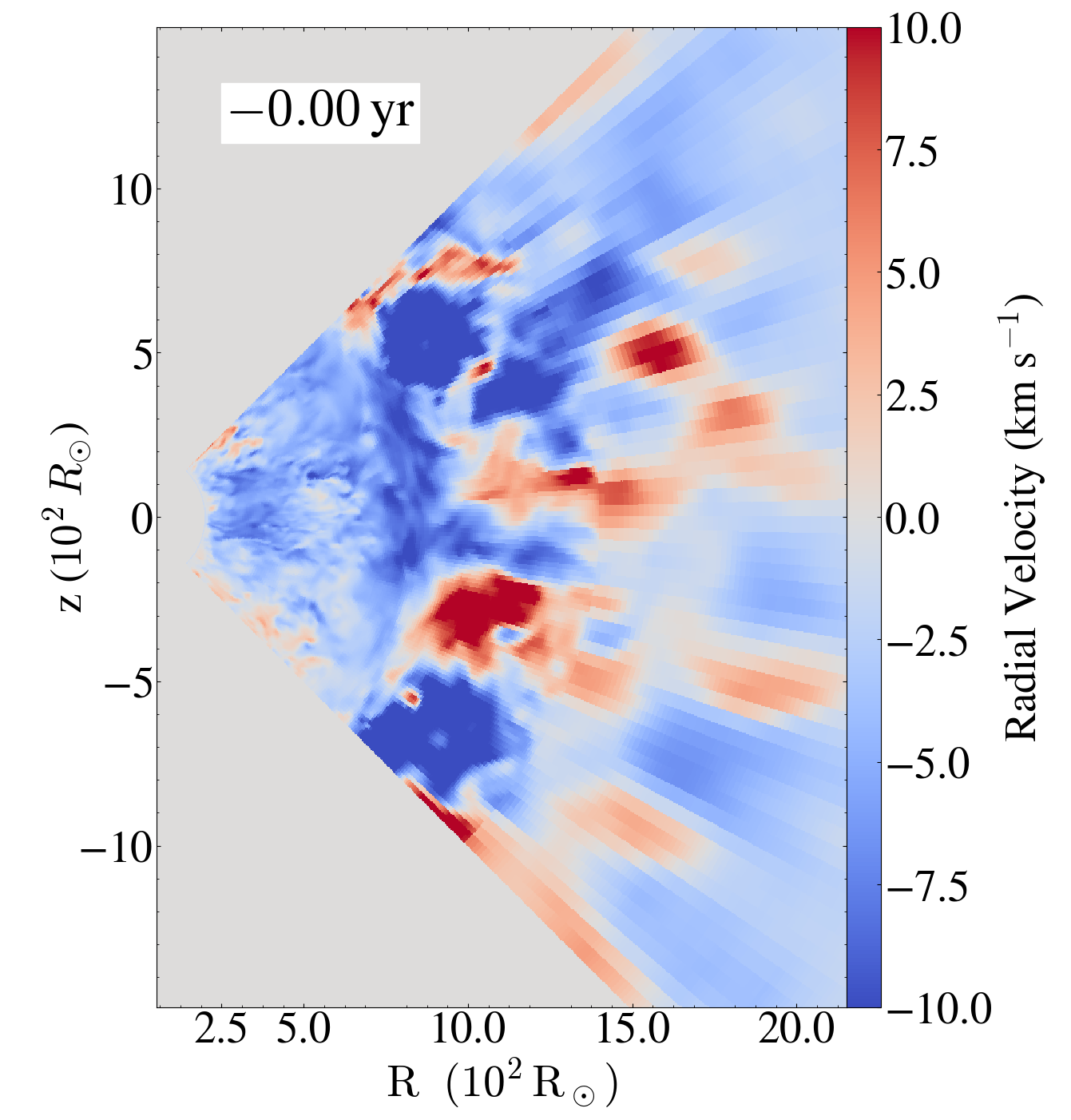}
\caption{Density and radial velocity slices at $\phi = \pi/2$, taken from the last snapshot of the convective equilibrium model before energy deposition. Convective structures are evident in the RSG envelope. 
} \label{fig:conv_eqm_slices}
\end{figure*}

\subsection{Convective Equilibrium Model}
\label{sec:conv_eqm}

To examine the 3D hydrodynamical response of a convective stellar envelope to energy deposition, we set up convection self-consistently by allowing our initial model envelope, mapped from the \mesa\ progenitor model, to relax for 25\,$t_{\rm dyn}$ ($\approx 6.5$\,years).
Due to the differences in numerical implementation (e.g., grid, EoS, and hydrodynamics solvers), the \texttt{FLASH} model initialized with the \texttt{MESA} profiles will not be in perfect equilibrium.
The model envelope pulsates slightly as it settles into a new quasi-steady equilibrium state. 
After $\approx$2\,years (7.6\,$t_{\rm dyn}$), convective motions seeded by numerical noise start developing.
Since our 3D implementation does not incorporate radiative losses near the envelope surface, convection is not actively driven. The convective motions in our model are the result of the 3D \texttt{FLASH} model envelope readjusting given the unstable entropy gradient.
As we will see, the velocity scale of convection is nonetheless comparable to the \texttt{MESA} model.

In Figure \ref{fig:E_time_prep} we show the time evolution of different components of energy as the initial model relaxes, develops convection, and establishes its new dynamical equilibrium.
The transient pulsations are most evident in the internal and the gravitational potential energy, which contribute to the majority of the total energy.
By the end of $25$\,$t_{\rm dyn}$, the model envelope fully settles into a quasi-steady state with global convective eddies. 
The total internal, gravitational, and kinetic energy are at $3.2 \times 10^{47}$\,erg, $-7.4 \times 10^{47}$\,erg, and $1.8 \times 10^{45}$\,erg, respectively. They sum to a total energy of $-4.2 \times 10^{47}$\,erg.
Kinetic energy is consistently small, amounting to only about 0.5\% of the total energy.

To insure that the convective equilibrium model represents a RSG envelope properly, we further inspect its envelope structure.
Figure \ref{fig:conv_eqm_profs} shows the angle-averaged radial profiles of gas density, temperature, thermal pressure (radiation plus gas), and velocities.
The color lines display the variations in the last 5\,$t_{\rm dyn}$ of the convective equilibrium evolution at 0.1\,$t_{\rm dyn}$ intervals.

Below $\approx$600\,$R_{\odot}$, the density, temperature, and pressure profiles from \mesa\ and \flash\ are almost identical. Gas velocities are about a few km\,s$^{-1}$ in this deeper layer, consistent with the convective velocity predicted by \mesa.
Above $\approx$720\,$R_{\odot}$ (stellar radius in the \mesa\ model), the convective equilibrium model has a low-density outer layer that extends to larger radii $\lesssim$1000\,$R_{\odot}$. This top layer is absent in the 1D \mesa\ model and is produced self-consistently in the 3D hydrodynamical evolution from the turbulent pressure of convection.
A similar `halo' layer was also observed in recent 3D radiation hydrodynamical models of RSG envelopes \citep{Goldberg21}.
The total mass above 720\,$R_{\odot}$ is 0.16\,$M_{\odot}$, 
about 2\% of the envelope mass. Upon energy injection, this low-density layer is the most susceptible to shock acceleration and mass ejection.
Figure \ref{fig:conv_eqm_slices} shows the azimuthal density and radial velocity slices of the convective equilibrium model.
It transitions from a relatively smooth density structure deep in the envelope to a highly aspherical outer layer, with convective structures visible throughout.

To aid further comparison, we repeat the above procedure for producing the 3D convective equilibrium model in 1D, i.e., by letting the initial \texttt{MESA} model relax for 25\,$t_{\rm dyn}$ and adjust to a quasi-steady state. The density profiles corresponding to this 1D control model are shown in the top left panel of Figure \ref{fig:conv_eqm_profs}. 
Since convection cannot manifest in 1D, the model envelope essentially maintains zero velocities everywhere except near the surface.
Adjustment to a new quasi-steady state causes the surface to pulsate slightly, leading to the formation of a similar low-density halo layer.
Density structures deeper in the envelope match closely with the \mesa\ initial model as in the 3D case.
Besides having the low-density surface layer, the 1D control model is quantitatively very similar to the static \texttt{MESA} envelope.

\begin{deluxetable*}{lccccccccc}[t]
\tablecaption{
Summary table of the parameters and results of the \flash\ hydrodynamical models.
\label{tab:hydro_results}}
\tablecolumns{8}
\tablewidth{0pt}
\tablehead{
\colhead{Model Name$\dagger$} & $E_{\rm bind}$ & $E_{\rm tot}$ & $E_{\rm dep} / E_{\rm bind}$ & $E_{\rm dep} / E_{\rm tot}$ & $\Delta t_{\rm dep}$ &  $M_{\rm ej}$ & $\Delta t_{\rm ej}$ & $\langle\dot{M}\rangle$ & KE$_{\rm ej}$ \\
 & ($-10^{47}$\,erg) & ($-10^{47}$\,erg) & & & (day) & ($M_{\odot}$) & (year) & ($M_{\odot}\,\textrm{yr}^{-1}$) & (10$^{46}\,{\rm erg}$)
}
\startdata
1DMESA0.5 & 7.67 & 4.34 & 0.28 & 0.5 & 20 & 0.15 & 1.43 & 0.11 & 0.55\\
1DCONV0.5 & 7.54 & 4.19 & 0.28 & 0.5 & 20 & 0.15 & 1.78 & 0.082 & 0.50 \\
1DCONV0.5LR & 7.54 & 4.19 & 0.28 & 0.5 & 40 & 0.063 & 1.74 & 0.036 & 0.15 \\
1DCONV0.5HR & 7.54 & 4.19 & 0.28 & 0.5 & 10 & 0.17 & 1.78 & 0.094 & 0.65 \\
\hline
1DCONV0.09 & 7.54 & 4.19 & 0.28 & 0.5 & 20 & 0.0 & 0.0 & 0.0 & 0.0 \\
1DCONV0.25 & 7.54 & 4.19 & 0.28 & 0.5 & 20 & 0.86 $\times 10^{-2}$ & 1.74 & 0.49 $\times 10^{-2}$ & 0.023 \\
1DCONV0.75 & 7.54 & 4.19 & 0.28 & 0.5 & 20 & 0.47 & 1.87 & 0.25 & 2.06 \\
1DCONV0.92 & 7.54 & 4.19 & 0.28 & 0.5 & 20 & 0.84 & 1.97 & 0.43 & 3.9 \\
\hline
\textbf{3DCONV0.5} & \textbf{7.42} & \textbf{4.16} & \textbf{0.28} & \textbf{0.5} & \textbf{20} & \textbf{0.50} & \textbf{2.38} & \textbf{0.21} & \textbf{1.72} \\
3DVTAN0.5 & 7.42 & 4.17 & 0.28 & 0.5 & 20 & 0.25 & 2.35 & 0.11 & 0.86 \\
3DVRAD0.5 & 7.42 & 4.17 & 0.28 & 0.5 & 20 & 0.48 & 2.35 & 0.20 & 1.72\\
3DAVGD0.5 & 7.41 & 4.15 & 0.28 & 0.5 & 20 & 0.47 & 2.31 & 0.21 & 1.58 \\
3DMESA0.5 & 7.67 & 4.34 & 0.28 & 0.5 & 20 & 0.15 & 1.43 & 0.11 & 0.55 \\
\hline
3DCONV0.09 & 7.42 & 4.16 & 0.05 & 0.09 & 20 & 2.95 $\times 10^{-4}$ & 3.33 & 0.89 $\times 10^{-4}$ & 5.47 $\times 10^{-4}$ \\
3DCONV0.25 & 7.42 & 4.16 & 0.14 & 0.25 & 20 & 0.038 & 1.97 & 0.019 & 0.10 \\
3DCONV0.75 & 7.42 & 4.16 & 0.42 & 0.75 & 20 & 1.52 & 2.60 & 0.58 & 6.51 \\
3DCONV0.92 & 7.42 & 4.16 & 0.52 & 0.92 & 20 & 2.32 & 2.76 & 0.84 & 11.3\\
\hline
3DCONV0.5LR & 7.42 & 4.16 & 0.28 & 0.25 & 40 & 0.36 & 2.28 & 0.16 & 1.07 \\
3DCONV0.5HR & 7.42 & 4.16 & 0.28 & 0.25 & 10 & 0.54 & 2.31 & 0.23 & 1.95 \\
\hline
\enddata
\tablenotetext{}{\textbf{Note:} Columns are defined as follows. (1) Model name indicating the dimension, type of pre-deposition RSG structure, and the amount of energy deposition, (2) amount of internal energy deposited relative to the total envelope energy, (3) duration of energy deposition, (5) time-averaged mass loss rate, (6) total mass of ejecta, (7) total kinetic energy of ejecta.}
\tablenotetext{}{\textbf{$\dagger$Model Nomenclature:} 3DCONV: energy deposition simulations based on the 3D convective equilibrium model (Section \ref{sec:conv_eqm}); 1DCONV: control runs started with the convective equilibrium model repeated in 1D; 3DVTAN (3DVRAD): same as 3DCONV, but with the radial (tangential) component of the convective velocity field zeroed out; 3DAVGD: same as 3DCONV, but with the angular fluctuations of density averaged out; MESA: model initialized directly from the hydrostatic \mesa\ profiles.}
\end{deluxetable*} 

Radiation contributes up to $\approx$40\% of pressure deep in the model envelope. In the deep interior, the \texttt{Helmholtz} EoS is a reliable description of the pressure contributions by both radiation and gas.
Above $\approx$850\,$R_{\odot}$, the assumed isotropic radiation pressure description of the EoS becomes inaccurate, as does the fully ionized assumption.
However, this radiation-dominated layer only has a total mass of $6.5 \times 10^{-3}$\,$M_{\odot}$. 

\hlm{The thermal timescale of the envelope is estimated to be $t_{\rm th} = |E_{\rm bind}| / L_{*} \approx 57.2$\,years, which is longer than the duration of the convective equilibrium model. Ideally, a proper simulation of the envelope would include radiation transport and cover longer time periods such as the thermal timescale, but it is computationally rather costly. 
With radiation transport enabled and given sufficient time, the envelope could in principle thermally relax and may change in structure. However, radiative cooling would act to maintain the entropy gradient and drive convection in the outer envelope. The convective equilibrium model has been evolved long enough to cover $\approx$20 convection turnover timescales, $t_{\rm conv} = \textrm{H}/v_{\rm conv} \approx 0.3\,$year, where $H = 50$\,$R_{\odot}$ and $v_{\rm conv} = 4$\,km\,s$^{-1}$ are the typical pressure scale height and convective velocity in the envelope.
In effect, we rely on the \texttt{MESA} model (which captures thermal effects) to obtain the general stellar structure, and use 3D hydrodynamics to better represent the structure of convective regions, which is more relevant in determining the outer envelope structure prior to core collapse. Our modeling approach should therefore provide a qualitatively reasonable representation of a convective RSG envelope.}
Overall, the convective equilibrium model reproduces in 3D the envelope structure of the \mesa\ model with a well-developed convective layer.

\subsection{Model Suite}
\label{sec:model_suite}

The \emph{convective equilibrium} model introduced above will be the starting point of a suite of hydrodynamical simulations designed to study the response of a convective stellar envelope to episodes of  energy deposition, a probable driving mechanism for pre-SN outbursts.
We focus on the case in which energy less than the binding energy is injected over a duration shorter than the dynamical time of the progenitor star $t_{\rm dyn} = \sqrt{R^{3}_{*}/GM_{*}}$ = 93.8\,days. 
We will explore how the results depend on the initial RSG envelope properties, dimensionality, as well as the amount and rate of energy deposition.

\begin{figure*}
\centering
\begin{tabular}{c c}
\textbf{\Large 3DCONV0.5} & \textbf{\Large 3DMESA0.5} \\
\includegraphics[height=9.5cm]{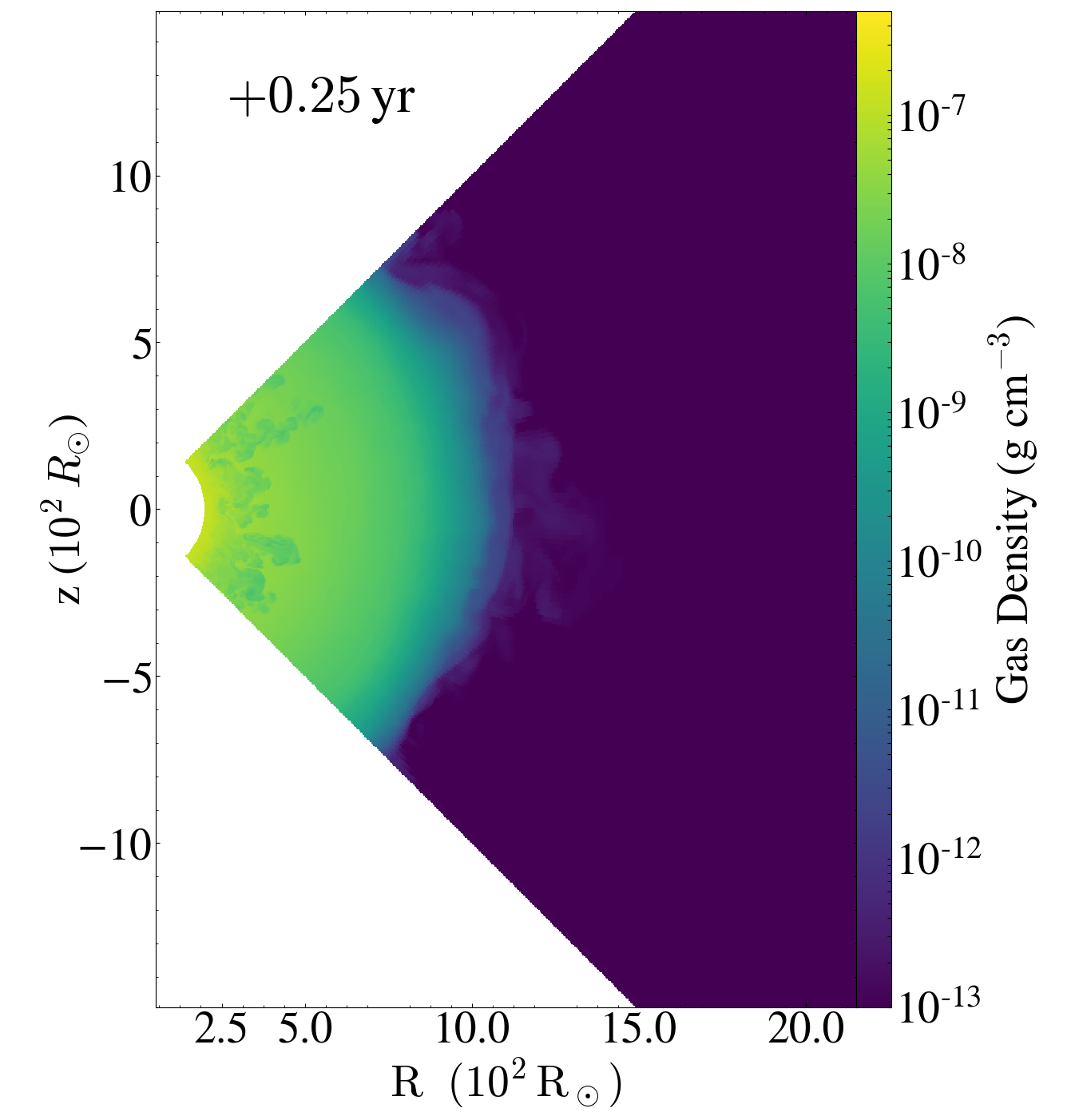}
&
{\includegraphics[height=9.5cm]{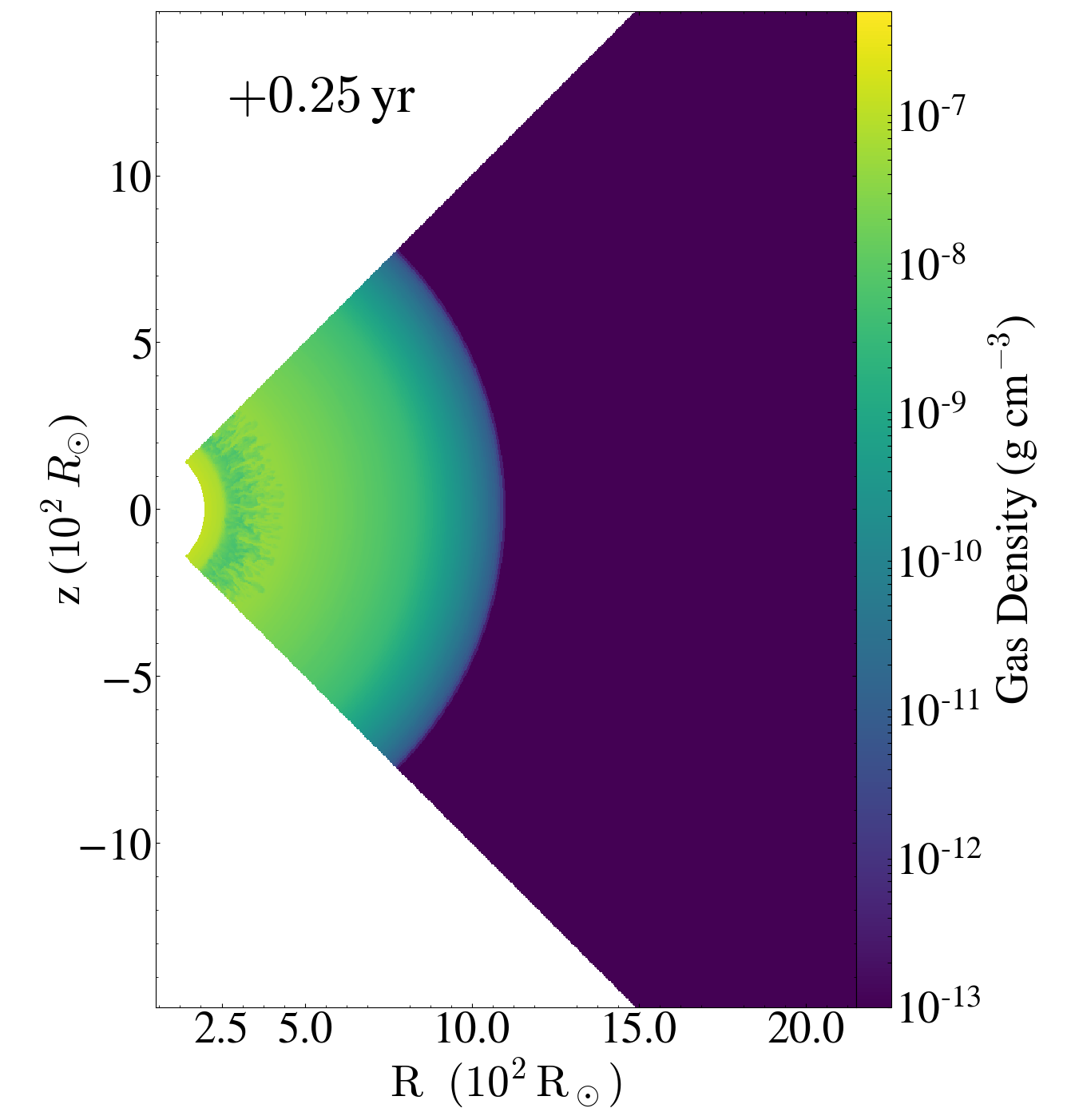}} \\
\includegraphics[height=9.5cm]{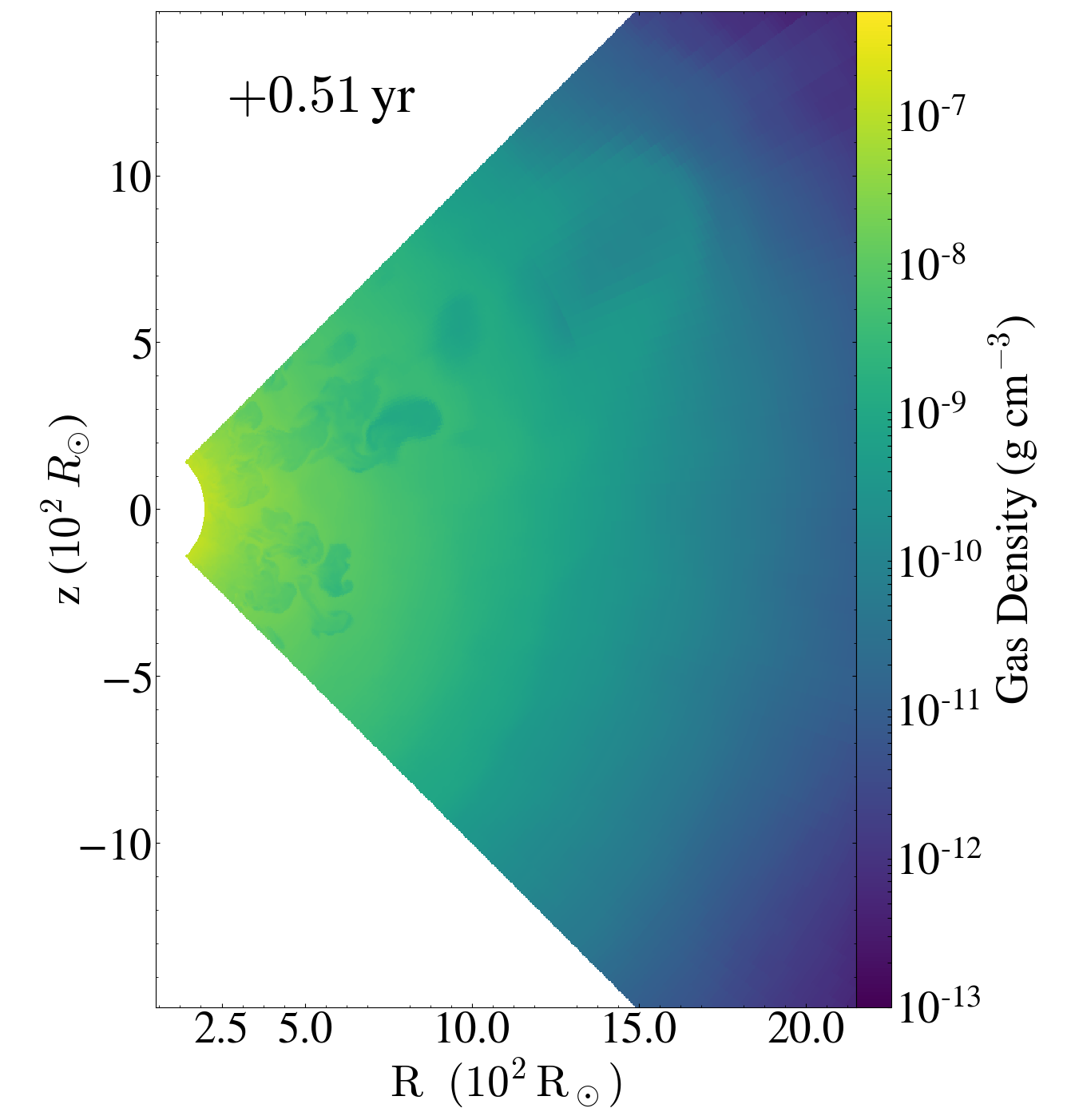}
&
\includegraphics[height=9.5cm]{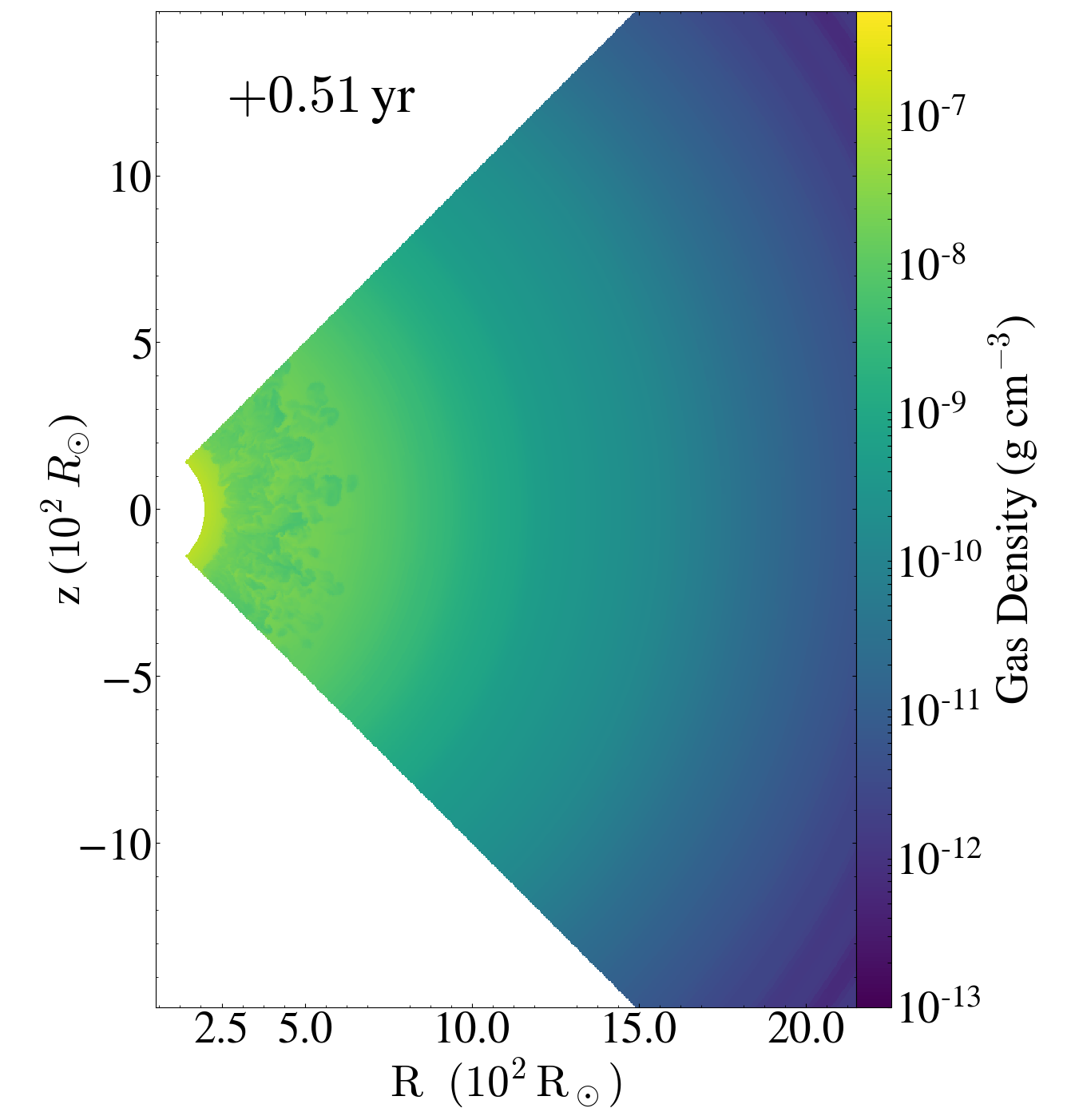}
\end{tabular}
\caption{Density slices of the 3DCONV0.5 fiducial (left) and 3DMESA0.5 (right) models at 1\,$t_{\rm dyn}$ (top) and 2\,$t_{\rm dyn}$ (bottom) after energy deposition. The timestamps display the time since the beginning of the energy injection. Non-uniform shock propagation in the 3DCONV0.5 model accelerates gas above escape velocity in some lines of sight, whereas in 3DMESA0.5 the envelope expands coherently despite the RTIs. 
} \label{fig:dens_slices_tile_1}
\end{figure*}

\begin{figure*}
\centering
\begin{tabular}{c c}
\textbf{\Large 3DCONV0.5} & \textbf{\Large 3DMESA0.5} \\
\includegraphics[height=9.5cm]{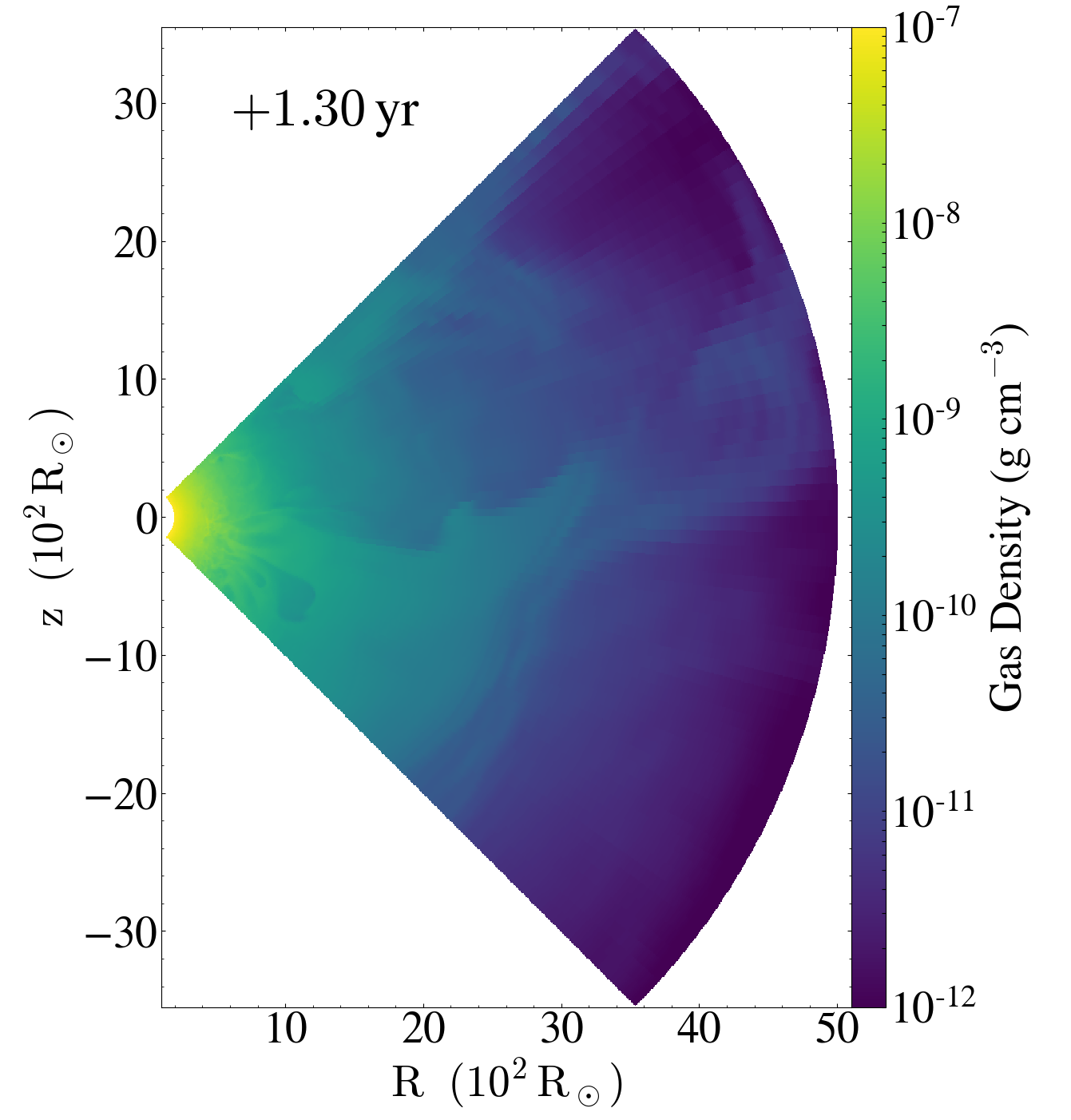}
&
\includegraphics[height=9.5cm]{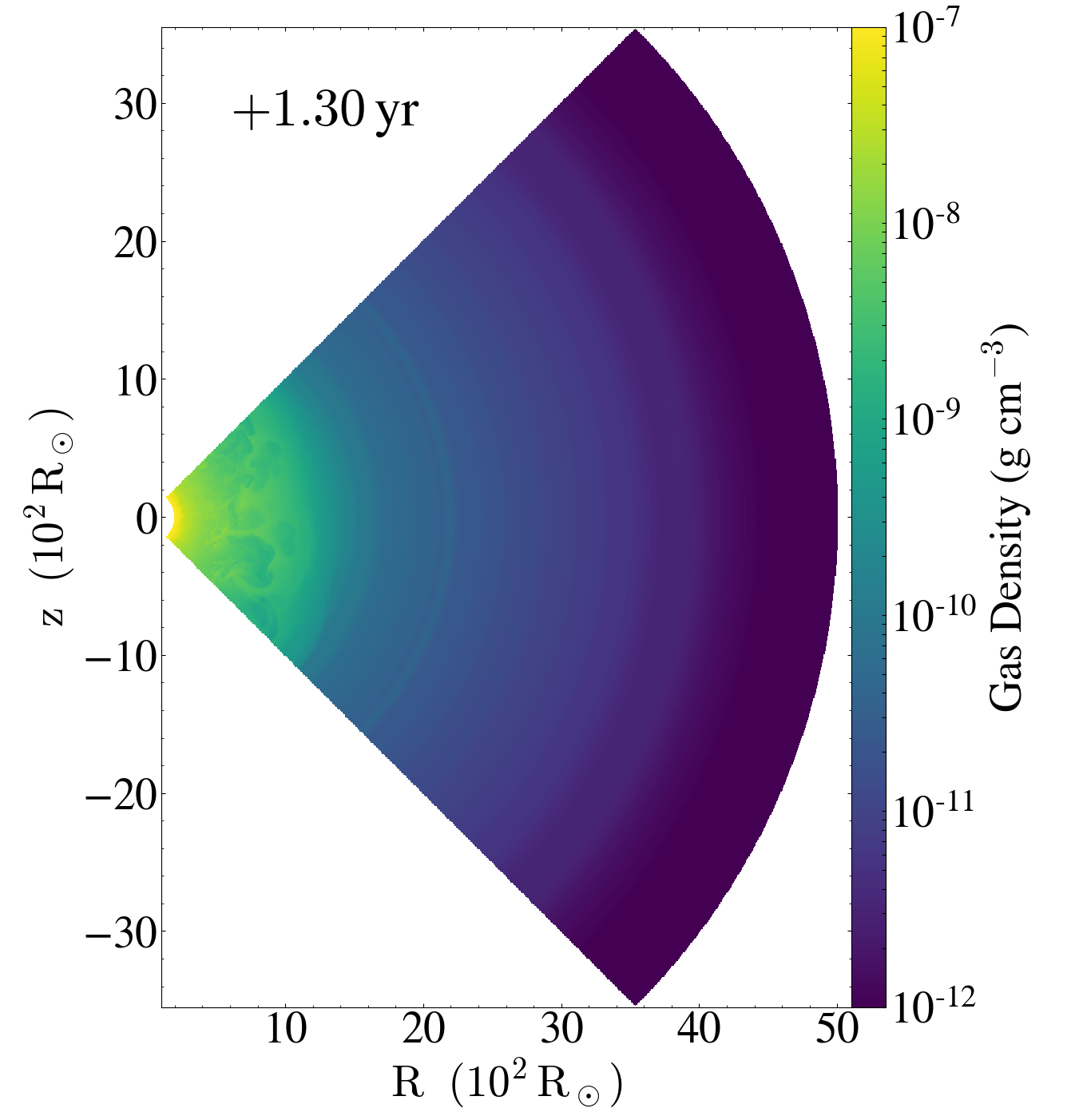} \\
\includegraphics[height=9.5cm]{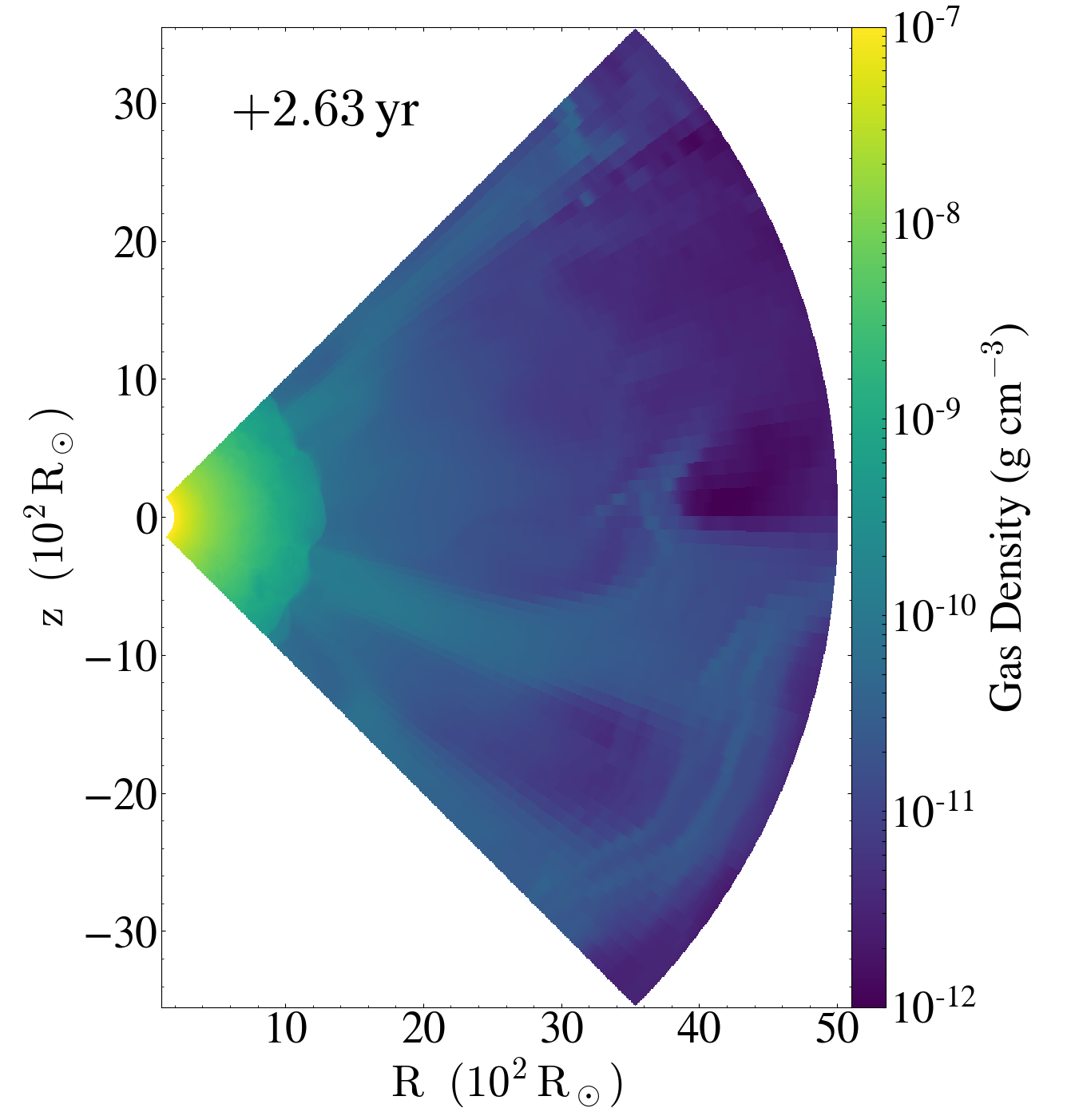}
&
\includegraphics[height=9.5cm]{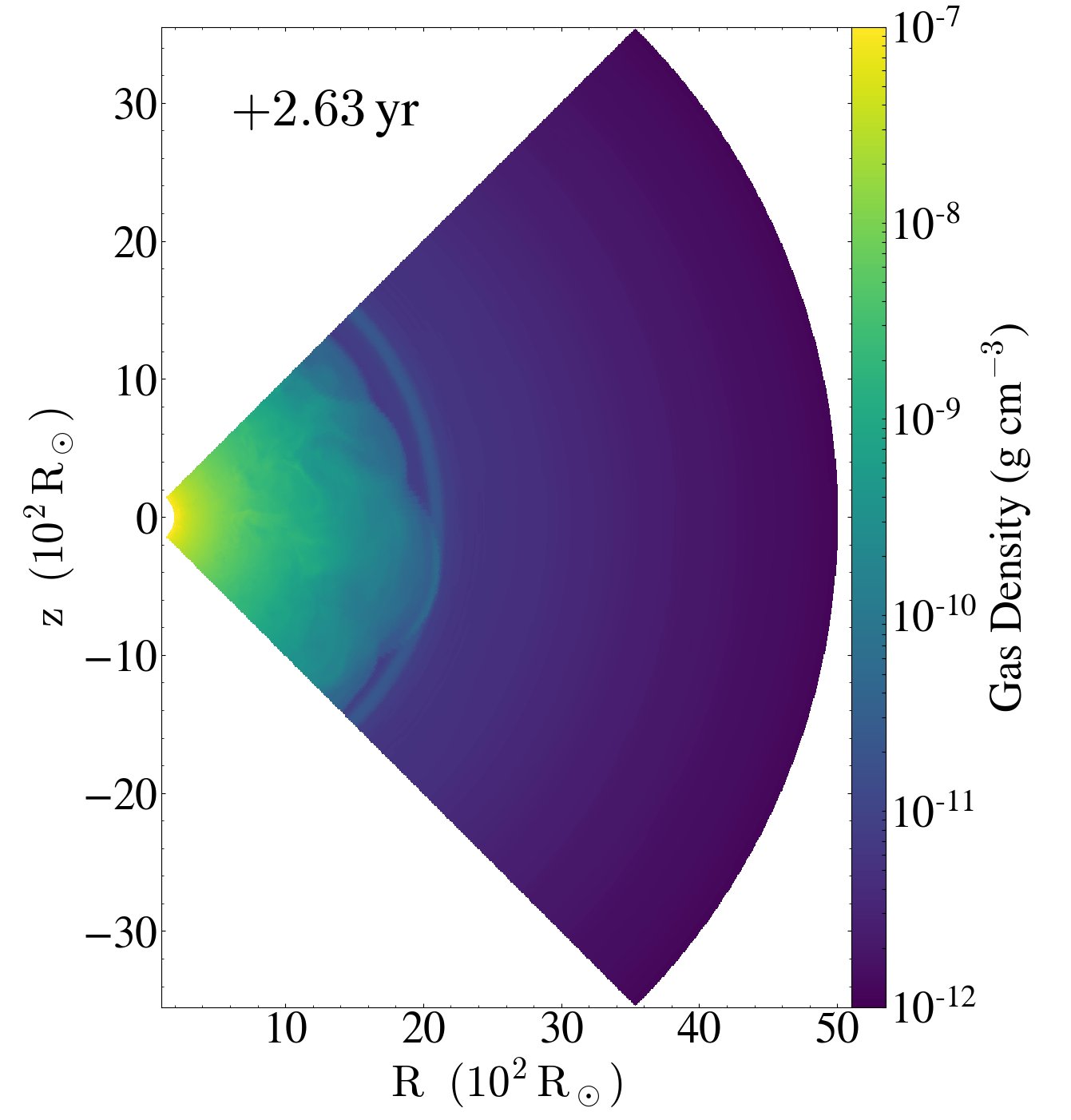}
\end{tabular}
\caption{Density slices of the 3DCONV0.5 fiducial (left) and 3DMESA0.5 (right) models at 5\,$t_{\rm dyn}$ (top) and 10\,$t_{\rm dyn}$ (bottom) after energy deposition.
Aspherical distributions of the ejected mass are apparent in 3DCONV0.5. In 3DMESA0.5, the envelope reaches a maximum radius of $\approx$1500\,$R_{\odot}$ and falls back into an inflated and turbulent state; unlike 3DCONV0.5; gas in the inner envelope cannot escape.
} \label{fig:dens_slices_tile_2}
\end{figure*}

Starting with the convective equilibrium model, the fiducial model 3DCONV0.5 focuses on the response of a realistic RSG envelope to an energy deposition over a short timescale.
We probe the non-terminal regime where the deposition will not lead to a supernova event or a complete ejection of the progenitor envelope.
Thermal energy equivalent to half of the envelope's total (internal $+$ kinetic $+$ gravitational) energy, which amounts to $2\times 10^{47}$\,erg, is injected within a spherical shell at radius $R_{\rm dep} = 240 \pm 10\,R_{\odot}$ near the base over 20\,days ($\approx$0.2\,$t_{\rm dyn}$).
The energy injection timescale of 20\,days is shorter than the dynamical time of the progenitor star as a whole. 
It is, however, longer than the local dynamical time at the site of the energy injection $t_{\rm dyn, dep} = \Delta R_{\rm dep}/c_{\rm s, dep}$ = 2.7\,days, where $c_{\rm s, dep} = 60$\,km\,s$^{-1}$ is the speed of sound at the deposition radius, allowing for that region to nearly hydrostatically adjust to the energy injection.

The parameters used in the model suite are summarized in Table \ref{tab:hydro_results}.
The choices of the energy budget and deposition timescale are informed by the recent studies of \citet{WF21, WF22}, which concluded that wave energy dissipation can contribute $10^{46-47}$\,erg in the months to years prior to core collapse.
Our model setup differs from theirs in that energy is injected at larger radius than the $\sim10$\,$R_{\odot}$ predicted for heating by gravity wave deposition.

To compare with existing models in the literature and to expose potential 3D effects, we deposit energy in the same manner but instead initialize the model envelope directly with the hydrostatic \mesa\ density and temperature profiles, both in 1D (1DMESA0.5) and in 3D (3DMESA0.5).
Without the outer low-density layer developing from the initial relaxation, the 1D/3DMESA0.5 models are slightly more compact and bound (3\% higher gravitational binding energy). The key difference from the fiducial model is the absence of convection.

To reveal how the presence of convective motions in the RSG envelope manifests in post-deposition gas dynamics, we repeat the fiducial model with its radial and tangential components of the velocity field zeroed out, in model 3DVTAN0.5 and 3DVRAD0.5, respectively.
In 3DAVGD0.5, we smooth out the density structures in angle to investigate the role of pre-existing density fluctuations in the envelope, keeping the velocity field unaltered.

\begin{figure}
\includegraphics[width=\columnwidth]{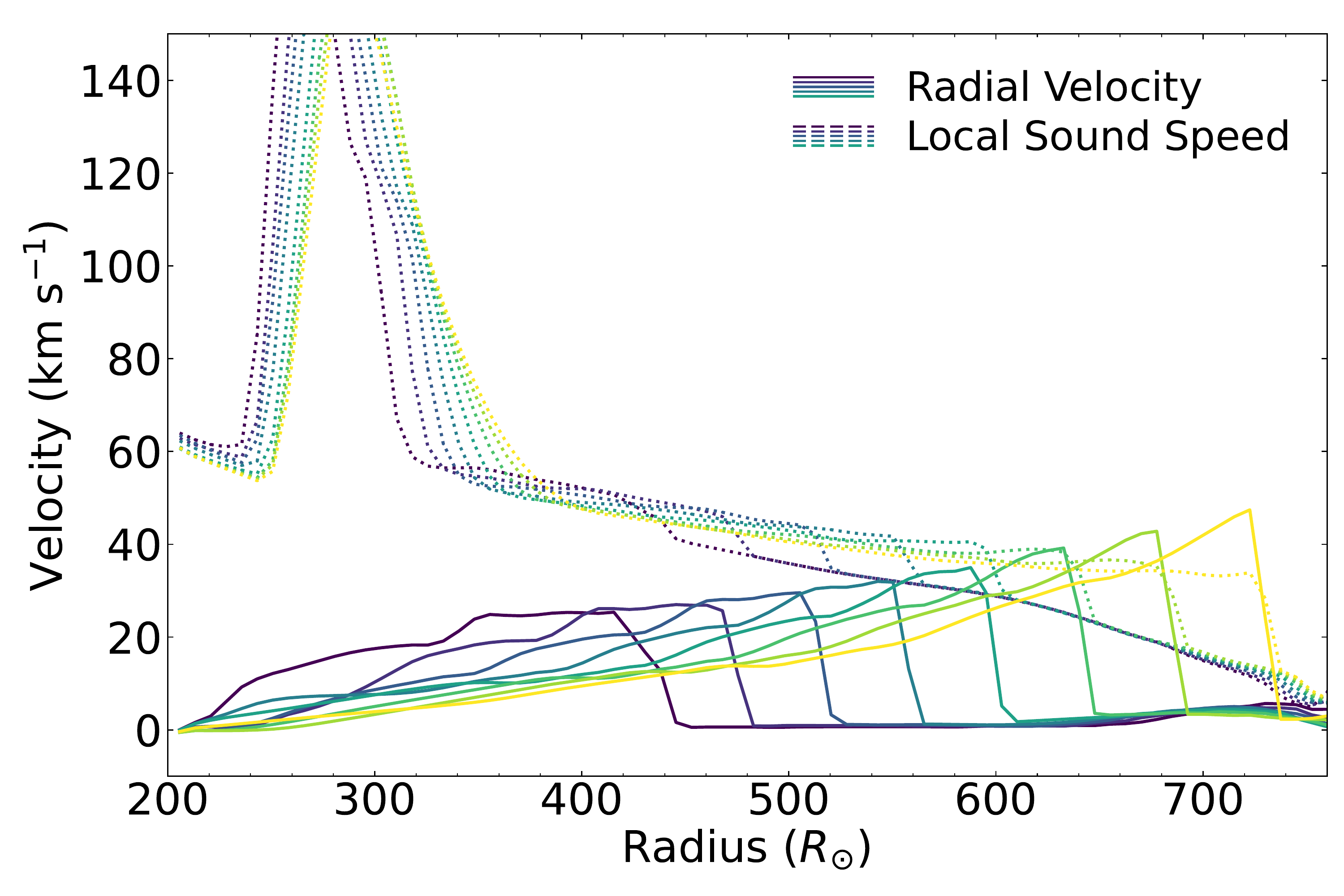}
\caption{Radial velocity profiles of the pulse launched by energy deposition from the 3DMESA0.5 run. The curves show the evolution from 0.3\,$t_{\rm dyn}$ to 0.75\,$t_{\rm dyn}$ at intervals of 0.06\,$t_{\rm dyn}$. The pulse strengthens and becomes a shock at the sonic radius of 640\,$R_{\odot}$. The peaks in the speed of sound near $\approx$300\,$R_{\odot}$ are the results of the energy deposition.
} \label{fig:shock_form}
\end{figure}

We further explore the dependence on the amount of energy deposited in the model series 3DCONV0.09, 3DCONV0.25, 3DCONV0.75, and 3DCONV0.9.
In 3DCONV0.5LR (low rate) and 3DCONV0.5HR (high rate), we analyze how the rate of energy deposition affect the results by halving and doubling the deposition rate in the fiducial model.
The 1DCONV[0.09-0.92] models differ from 1DMESA0.5 by first allowing the initial \mesa\ model to relax for 25\,$t_{\rm dyn}$ before energy deposition, same procedure as when the 3D convective equilibrium is prepared (see Section \ref{sec:conv_eqm}).
We recognize that it is a misnomer to denote 1D models with `CONV' since convection cannot develop in 1D. The naming is to allow convenient reference to the 1D/3D model runs with otherwise identical initialization procedures.

\section{Results}
\label{sec:results}

In this section, we describe in detail the results from the hydrodynamical models.
We first focus on the crucial importance of including convection when modeling mass loss from pre-SN outburst (Section \ref{sec:conv_role}).
We then compare the models in 1D and 3D to highlight the model dependence on dimensionality (Section \ref{sec:1D3D}).
We study trends of varying the amount of energy deposited and the deposition rate (Section \ref{sec:dep_study}).
In Section \ref{sec:3dejecta}, we investigate the 3D distribution of ejected mass and the properties of the resultant CSM environment.

\subsection{The Importance of Convective Initial Conditions}
\label{sec:conv_role}

We first compare the hydrodynamics of the two key 3D models of this work, namely, the convective equilibrium model (3DCONV0.5) and its static counterpart initialized directly from the \mesa\ model envelope (3DMESA0.5).
We display the density slices from the two models taken at $t_{\rm dyn}$, 2\,$t_{\rm dyn}$, 5\,$t_{\rm dyn}$, and 10\,$t_{\rm dyn}$ since the start of energy deposition in Figure \ref{fig:dens_slices_tile_1} and \ref{fig:dens_slices_tile_2}.

Shortly after energy deposition begins, heating creates an initially narrow high-pressure zone near the base of the envelope.
{Since the energy injection timescale is longer than the dynamical time at the injection zone, the heated layer initially expands hydrostatically. The radial expansion drives an outward pressure pulse. As the pulse travels outward, the local speed of sound decreases and the pulse strengthens into a shock as it sweeps through the envelope and approaches the surface, similar to the dynamics described by \citet{Coughlin18} and \citet{Linial21}.
As shown in Figure \ref{fig:shock_form} for the 3DMESA0.5 run, the outward-moving pulse becomes sonic at $\approx$640\,$R_{\odot}$. About 0.2\,$M_{\odot}$ of gas is located above this sonic radius, which provides an upper limit for the shock-induced mass loss.
The angle-averaged shock dynamics is very similar between the 3DCONV0.5 and 3DMESA0.5 models at this beginning stage.}
However, the detailed 3D gas dynamics differs significantly, resulting in substantially different amounts of final mass loss.

\begin{figure}
\includegraphics[width=\columnwidth]{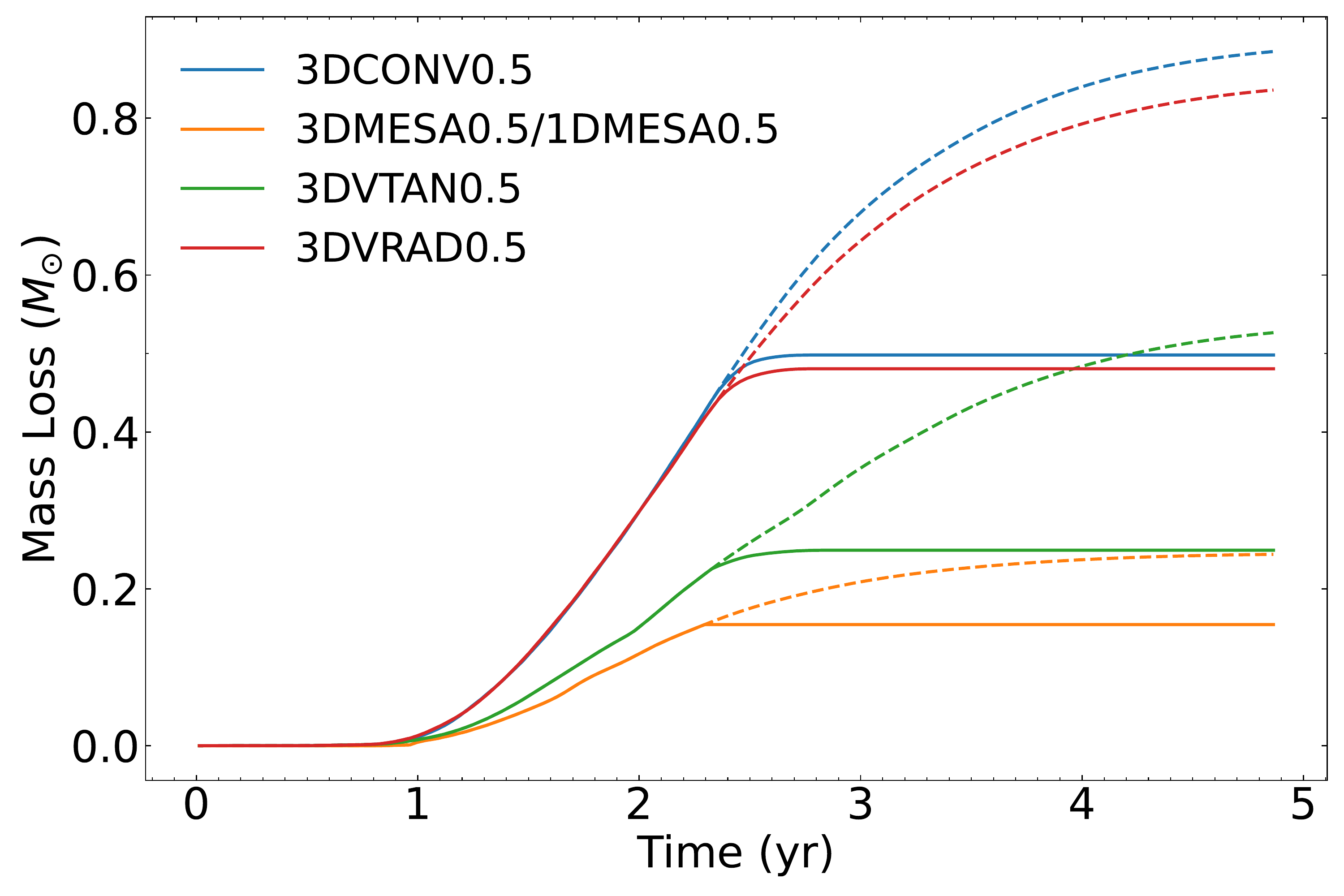}
\caption{Cumulative mass loss of the RSG outburst models with different initial envelope structures. Solid lines track the mass expected to be ejected (with radial velocity exceeding the escape velocity); dashed lines simply tally mass leaving the outer simulation boundary regardless of velocity.
Model envelopes with initial radial motions (3DCONV0.5 and 3DVRAD0.5) are susceptible to non-uniform shock propagation, resulting in considerably higher mass loss. With the initial radial velocities removed, 3DVTAN0.5 behaves qualitatively similar to the static models 3D/1DMESA0.5. 
} \label{fig:mass_loss}
\end{figure}

\begin{figure}
\includegraphics[width=\columnwidth]{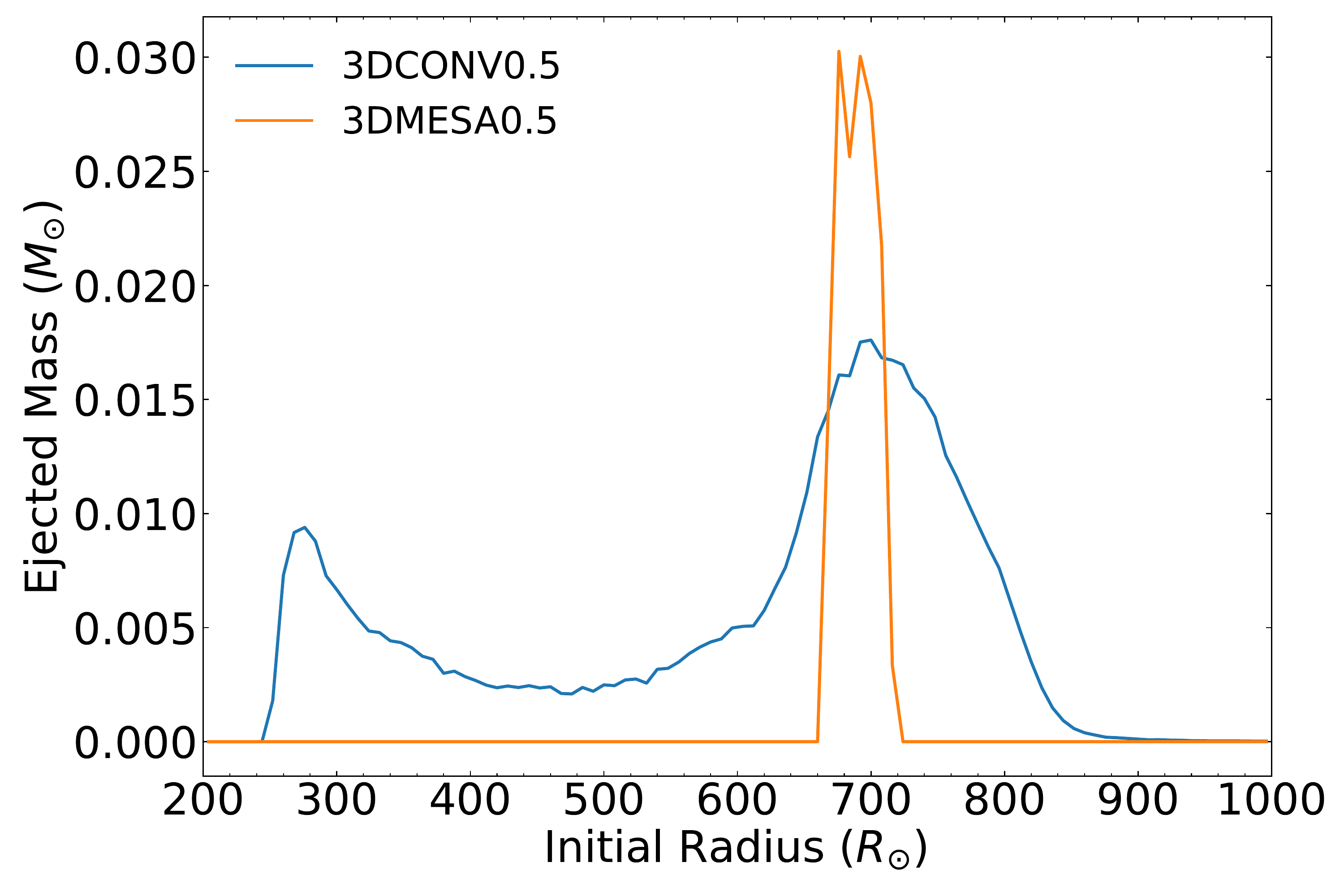}
\caption{Pre-deposition radius histogram of the ejected mass from the 3DCONV0.5 and 3DMESA0.5 runs. Without convective initial conditions, only mass above the sonic radius can be ejected (3DMESA0.5). In 3DCONV0.5, mass down to the energy injection radius is able to escape through low-density channels set up by the convective velocities.  
} \label{fig:mass_loss_histogram}
\end{figure}

We show the cumulative mass loss in the two models in Figure \ref{fig:mass_loss}. The solid lines depict the amount of mass expected to be ejected from the envelope, i.e., mass that attains or exceeds the escape speed at the outer boundary, the \emph{ejecta mass} $M_{\rm ej}$. 
The dotted lines simply tally the mass flux across the outer boundary regardless of gas velocity. A total of $M_{\rm ej} = 0.5\,M_{\odot}$ is unbound in the 3DCONV0.5 model, whereas 3DMESA0.5 yields only $M_{\rm ej} = 0.15\,M_{\odot}$.
We further dissect the difference by showing the distributions of the pre-deposition radius of the ejected mass in Figure \ref{fig:mass_loss_histogram}, as tracked by a mass scalar quantity in \texttt{FLASH}.
In 3DMESA0.5, only mass above $\approx$660\,$R_{\odot}$ is ejected, which is consistent with the location and the mass on top of the sonic radius estimated above (see Figure \ref{fig:shock_form}).
In 3DCONV0.5, \emph{instabilities seeded by the initial convective motions carve out low-density channels through which gas from as deep as the energy injection zone is able to escape, increasing the total amount of mass ejected as a result of energy deposition.}

In order to understand the significant difference in $M_{\rm ej}$, we overplot the results of two control runs in Figure \ref{fig:mass_loss}. Model 3DVTAN0.5 (3DVRAD0.5) is identical to 3DCONV0.5 except that the radial (tangential) component of the convective velocity field is zeroed out. The cumulative mass loss from 3DVRAD0.5 is consistent with 3DCONV0.5, while 3DVTAN0.5 is similar to 3DMESA0.5.
In essence, the radial motion of convection readily provides outward-moving lines of sight through which shocks from energy deposition can more easily accelerate gas to high velocities. This effect is self-reinforcing -- gas that receives stronger acceleration can reach larger radii, creating channels for the high-pressure gas to preferably expand into. By 2\,$t_{\rm dyn}$, such bubbles can be seen to have punctured the envelope in multiple lines of sight, reaching radius $\gtrsim$1000\,$R_{\odot}$ and with radial velocity exceeding the local escape velocity (Figure \ref{fig:dens_slices_tile_1}, bottom left; Figure \ref{fig:dens_slices_tile_2}, top left).

\begin{figure*}
\centering
\begin{tabular}{c c}
    \textbf{\Large 3DCONV0.5} & \textbf{\Large 3DMESA0.5} \\
    \includegraphics[width=0.49\textwidth]{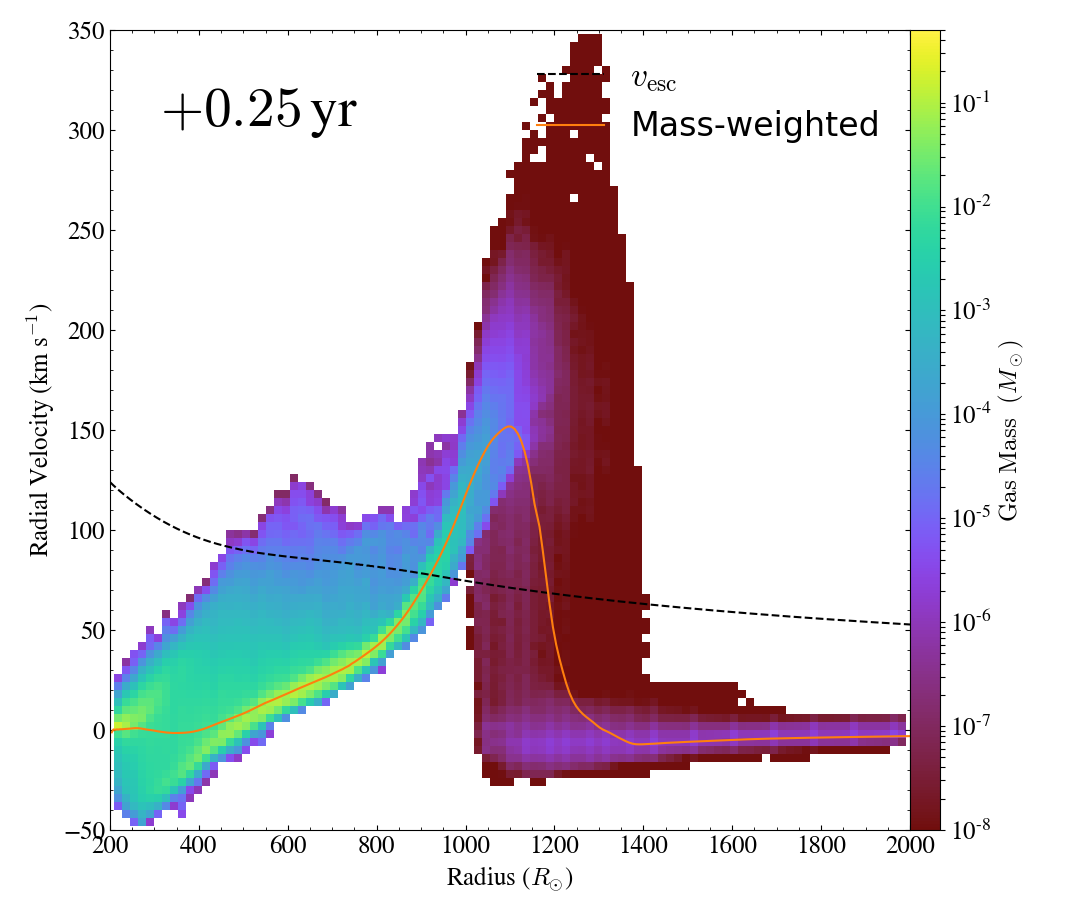} & 
    \includegraphics[width=0.49\textwidth]{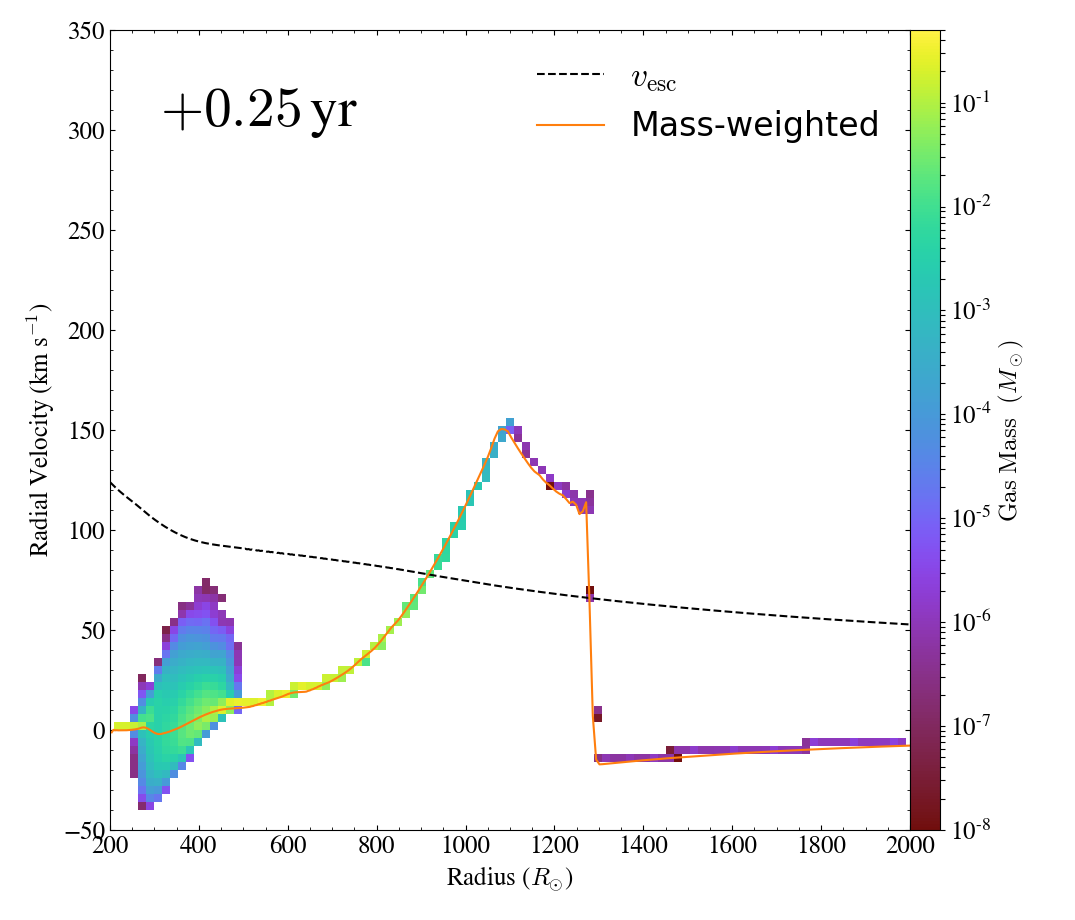} \\
    \includegraphics[width=0.49\textwidth]{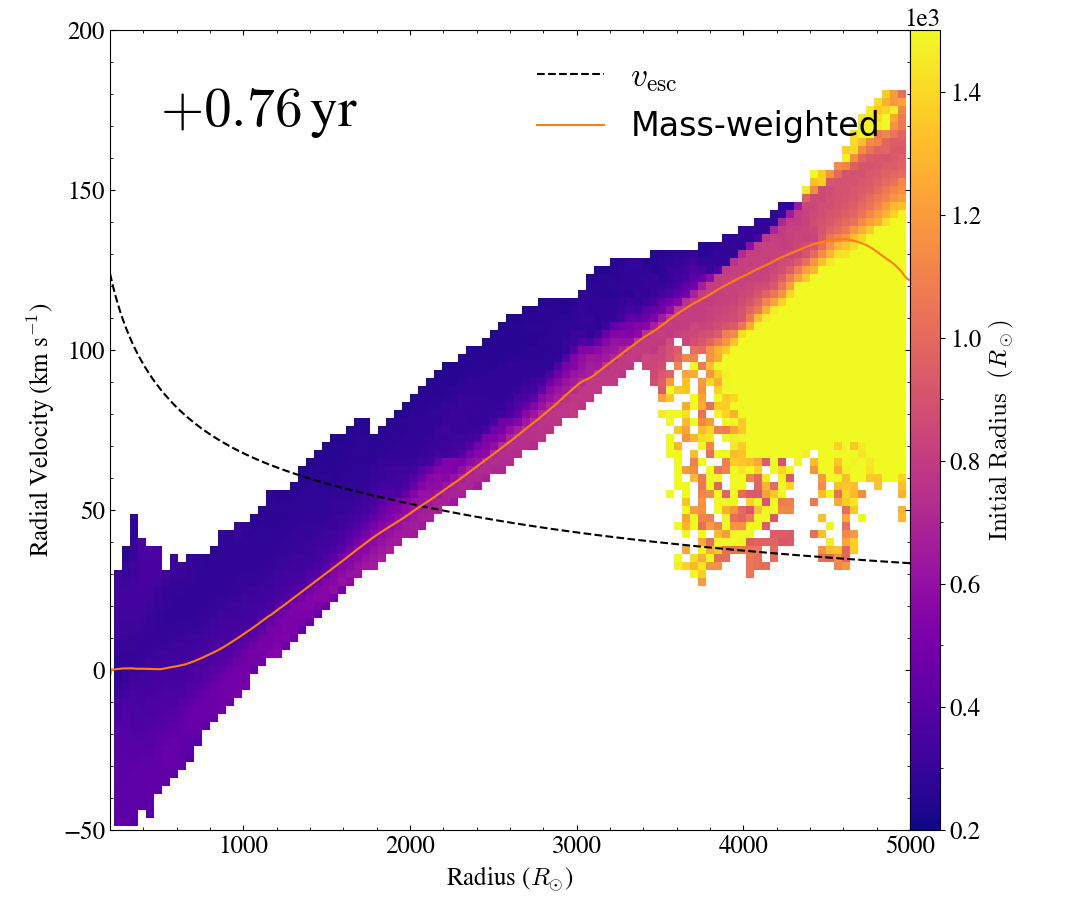} &
    \includegraphics[width=0.49\textwidth]{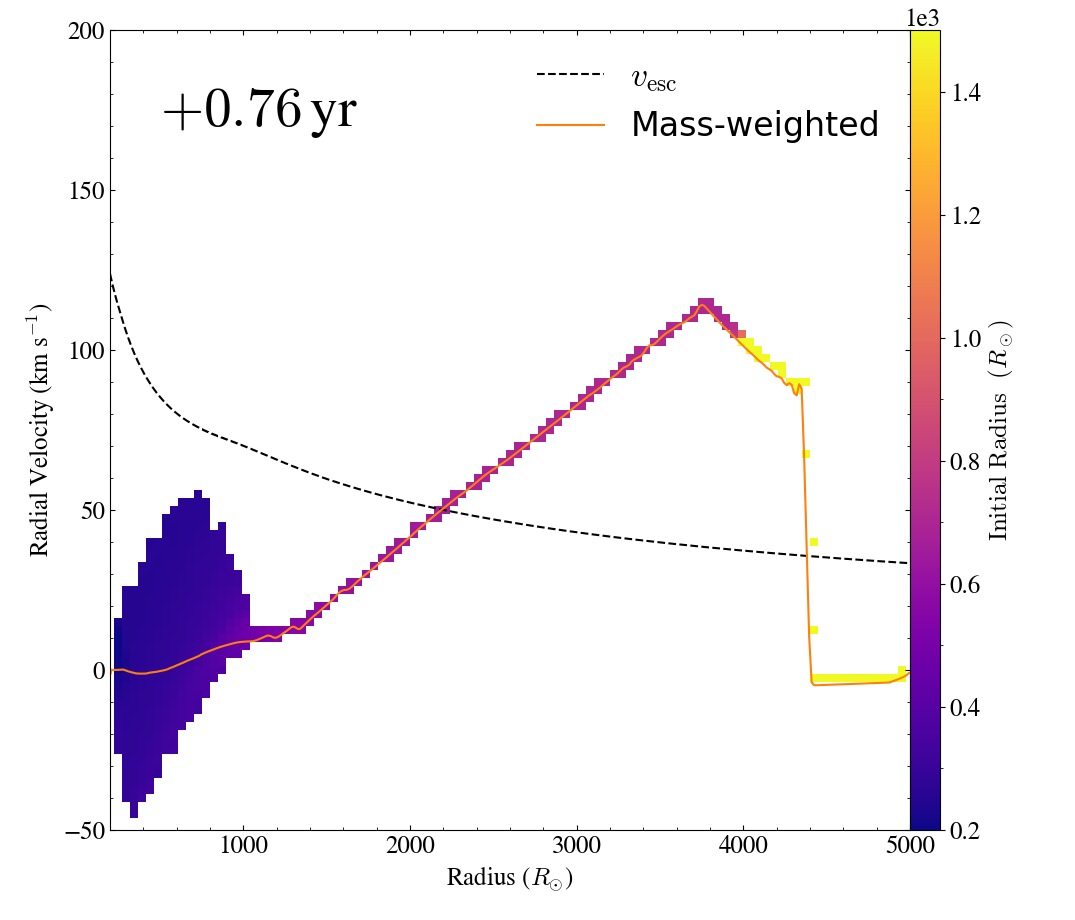}
\end{tabular}
\caption{Phase diagrams showing the distribution of gas mass (top) and their initial radii (bottom) on the radial velocity-radius plane from the 3DCONV0.5 (left) and 3DMESA0.5 (right) runs. The diagrams are taken at $t = t_{\rm dyn}$ and 3\,$t_{\rm dyn}$. 
The mass-weighted radial velocity (orange lines) and local escape velocity (black dashed lines) are shown for reference. 
The non-uniform shock propagation in the convective envelope of 3DCONV0.5 causes a much wider spread of velocity and allows gas from the deep interior to be accelerated above the escape velocity (dark blue band in the bottom left panel), leading to $\approx$3$\times$ more ejected mass.
} \label{fig:phase_velx_mass}
\end{figure*}

In the top panels of Figure \ref{fig:phase_velx_mass}, we show the distributions of gas mass in the 3DCONV0.5 and 3DMESA0.5 runs on the radial velocity-radius phase plane, taken at $t = t_{\rm dyn}$. As a result of the initial shock propagation, the mass-weighted radial velocity profiles (orange curves) in both models acquire a similar monotonic increase to $\approx$150\,km\,s$^{-1}$ near the outer edge of the envelopes.
However, shock propagation through 3DCONV0.5's convective envelope leads to rapid growth of instabilities and a much wider velocity dispersion around the averaged profile, resulting in a larger portion of the envelope being accelerated above the escape velocity of $\approx$90\,km\,s$^{-1}$ (c.f. gas mass above the black dashed lines).
The bottom panels of Figure \ref{fig:phase_velx_mass} are the mass-weighted initial radius distributions of the gas on the same phase plane at $t = 3\,t_{\rm dyn}$. Most evidently at $1000\,R_{\odot} \lesssim R \lesssim 4000\,R_{\odot}$, the non-uniform gas acceleration in 3DCONV0.5 allows gas from $R \lesssim 300\,R_{\odot}$ to reach $\gtrsim 1000\,R_{\odot}$ along certain lines of sight. Such gas dynamics is absent in the initially static 3DMESA0.5 model, which explains the significant difference in the final mass loss between the two models.

In contrast, behind the expanding envelope of 3DMESA0.5, RTIs generate finger-like structures with radial velocities of $\pm 50$\,km\,s$^{-1}$. Besides the RTI-induced density/velocity variations, the envelope expands mostly coherently above $R \gtrsim 1000\,R_{\odot}$. 
In the expanding envelope, gas mass closely traces the mass-weighted velocity profile.
At 5\,$t_{\rm dyn}$, the coherent shell-like structure in 3DMESA0.5 begins to break apart as the expanding envelope decelerates and falls back in certain directions (Figure \ref{fig:dens_slices_tile_2}, top right).
The bulk of the envelope reaches a maximum radial extent of $\approx$1500\,$R_{\odot}$ and cannot escape.
In both models, gas along under-accelerated directions eventually falls back and collides with the leftover envelope, leaving behind a turbulent and inflated envelope. 

At about $t = t_{\rm dyn}$, the shock induced by energy deposition reaches the envelope surface at velocities of $v_{\rm shock} \gtrsim$100--200\,km\,s$^{-1}$. Assuming a constant electron scattering opacity of $\kappa_{\rm es} = 0.33$\,cm$^{2}$\,g$^{-1}$ (corresponding to the hydrogen fraction of $X = 0.67$ from the \texttt{MESA} model), the optical depth at which radiation can diffuse ahead of the shock is $\tau_{\rm crit} = c / v_{\rm shock} \approx 5.6 \times 10^{3}$.
This critical optical depth is located at $R \approx 850$\,$R_{\odot}$ at $t = t_{\rm dyn}$, above which the envelope contains about $0.3$\,$M_{\odot}$.
The leakage of radiation at this depth could alter subsequent gas dynamics and potentially modify the amount of mass loss.
We encourage future radiation hydrodynamical calculations to verify the extent of mass loss from this radiative, partially ionized layer of the envelope.
Nevertheless, the non-uniform shock propagation in the convective envelope makes it possible for gas from deeper in the envelope to escape and causes a $\approx$3$\times$ increase in ejected mass (dark blue band in the bottom left panel of Figure \ref{fig:phase_velx_mass}).
Given that the difference in mass loss with and without initial convective velocities is comparable to the mass in the radiative layer,
we conclude that the effect of the convective velocities on the estimated mass loss is robust, and that convection can be an important consideration in modeling energy deposition in RSG envelopes in the non-terminal, short timescale limit.

\begin{figure}
\includegraphics[width=\columnwidth]{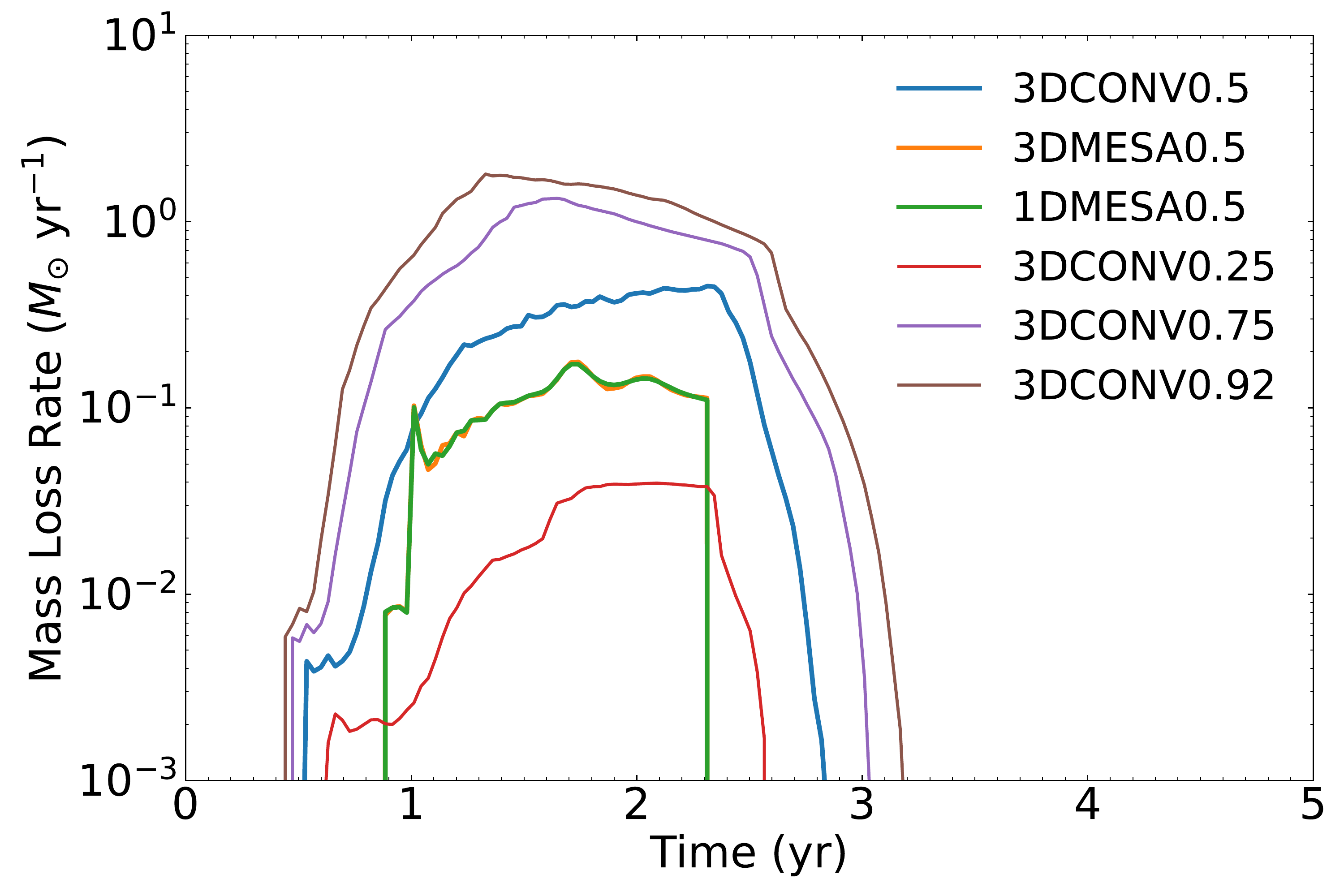}
\caption{Mass loss rate evolution of models with different initial structures and amounts of injected energy. The fiducial model's mass loss rate is a factor of few higher than 3DMESA0.5 throughout their entire mass loss episodes. 
} \label{fig:mass_loss_rate}
\end{figure}

Figure \ref{fig:mass_loss_rate} shows the mass loss rates of the two models. 
During its mass loss episode, the mass loss rate of the 3DCONV0.5 model rises rather sharply from $\approx$$10^{-3}$\,$M_{\odot}$\,yr$^{-1}$ to $\approx$$10^{-1}$\,$M_{\odot}$\,yr$^{-1}$ in about 0.5\,yr, after that it gradually increases to a peak of 0.5\,$M_{\odot}$\,yr$^{-1}$. The mass loss rate of 3DMESA0.5 is consistently a factor of few lower, exhibiting a more abrupt beginning and end, with a similar gradual increase in between.
We summarize $M_{\rm ej}$ in Table \ref{tab:hydro_results}. The time-averaged mass loss rate $\langle \dot{M} \rangle$ is also listed, defined as $\langle \dot{M} \rangle = M_{\rm ej} / \Delta t_{\rm ej}$, where $\Delta t_{\rm ej}$ is the time span in which the mass loss rate is positive.
The density fluctuations present in the initial convective equilibrium model play a minor role in determining post-deposition gas dynamics and mass loss. In the control run in which density variations in angle were smoothed out (3DAVGD0.5), we recover results that are almost identical to the fiducial model.

\begin{figure}
\includegraphics[width=\columnwidth]{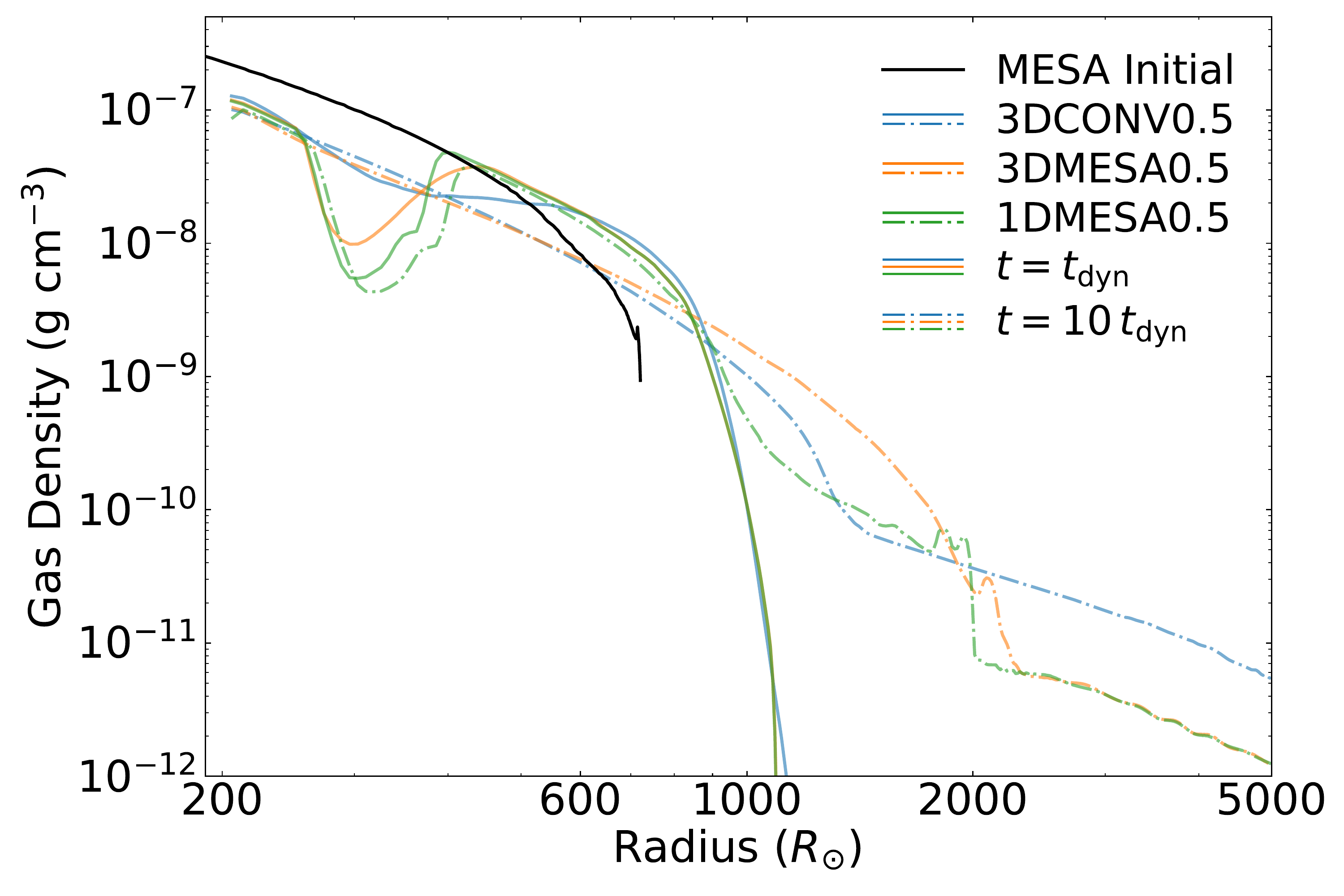}
\caption{Density profiles at two characteristic times $t = t_{\rm dyn}$ and $10\,t_{\rm dyn}$ from the 3DCONV0.5, 3DMESA0.5, and 1DMESA0.5 runs. Density inversion following energy deposition only manifests in the initially static models without convection (1D/3DMESA0.5). It persists in 1D (green dot-dashed line) but is later smoothed out by RTIs in 3D (orange dot-dashed line). Note the logarithmic scale in radius. 
} \label{fig:dens_prof_1D3D}
\end{figure}

\subsection{Comparing 1D and 3D Models}
\label{sec:1D3D}
We repeat model 3DMESA0.5 in 1D to reveal the fundamental effects of extending pre-SN outburst models in dimensions. 
In Figure \ref{fig:dens_prof_1D3D}, we compare the volume-weighted density profiles from 3DCONV0.5, 3DMESA0.5, and 1DMESA0.5 at two characteristic times.
At $t_{\rm dyn}$, the angle-averaged density distributions of the shocked envelope at radius $\gtrsim$500\,$R_{\odot}$ are comparable between models. Below 400\,$R_{\odot}$, model 1DMESA0.5 and 3DMESA0.5 both develop a density inversion, although the 3D case's is slightly more smoothed out due to early growth of RTIs. In model 3DCONV0.5 with intrinsic convection, no density inversion is observed.
Later at 10\,$t_{\rm dyn}$, the density inversion in 3DMESA0.5 is completely dissipated by the RTI-induced turbulence and its density profile in the interior matches well with 3DCONV0.5 (blue and orange dot-dashed lines). In the 1D case, the density inversion is persistent and unphysical. 
In the outer region $\gtrsim$2000\,$R_{\odot}$ of the 3DCONV0.5 run, the non-uniform gas acceleration along different lines of sight channels more gas from the inner envelope, giving rise to a more extended mass distribution with $\approx$10$\times$ higher average density.

We remark that the mass loss histories are identical between 1DMESA0.5 and 3DMESA0.5. It suggests that by extending from 1D to 3D, one can much more realistically capture the density evolution in the envelope in the presence of instabilities and avoid the unphysical density inversion. Without including convective motions in the envelope, the dynamics of the outer envelope where mass is unbound is effectively 1D and the $M_{\rm ej}$ estimates can be misleading.
Density profiles from 1DCONV0.5 are almost identical to 1DMESA0.5 and are therefore not shown.
The gravitational binding energy of 1DCONV0.5 before the energy deposition is only marginally ($\lesssim$2\%) lower than 1DMESA0.5, both models therefore result in almost identical $M_{\rm ej}$.

\subsection{Dependence on the Amount and Rates of Energy Deposition}
\label{sec:dep_study}

\begin{figure}
\centering
\includegraphics[width=\columnwidth]{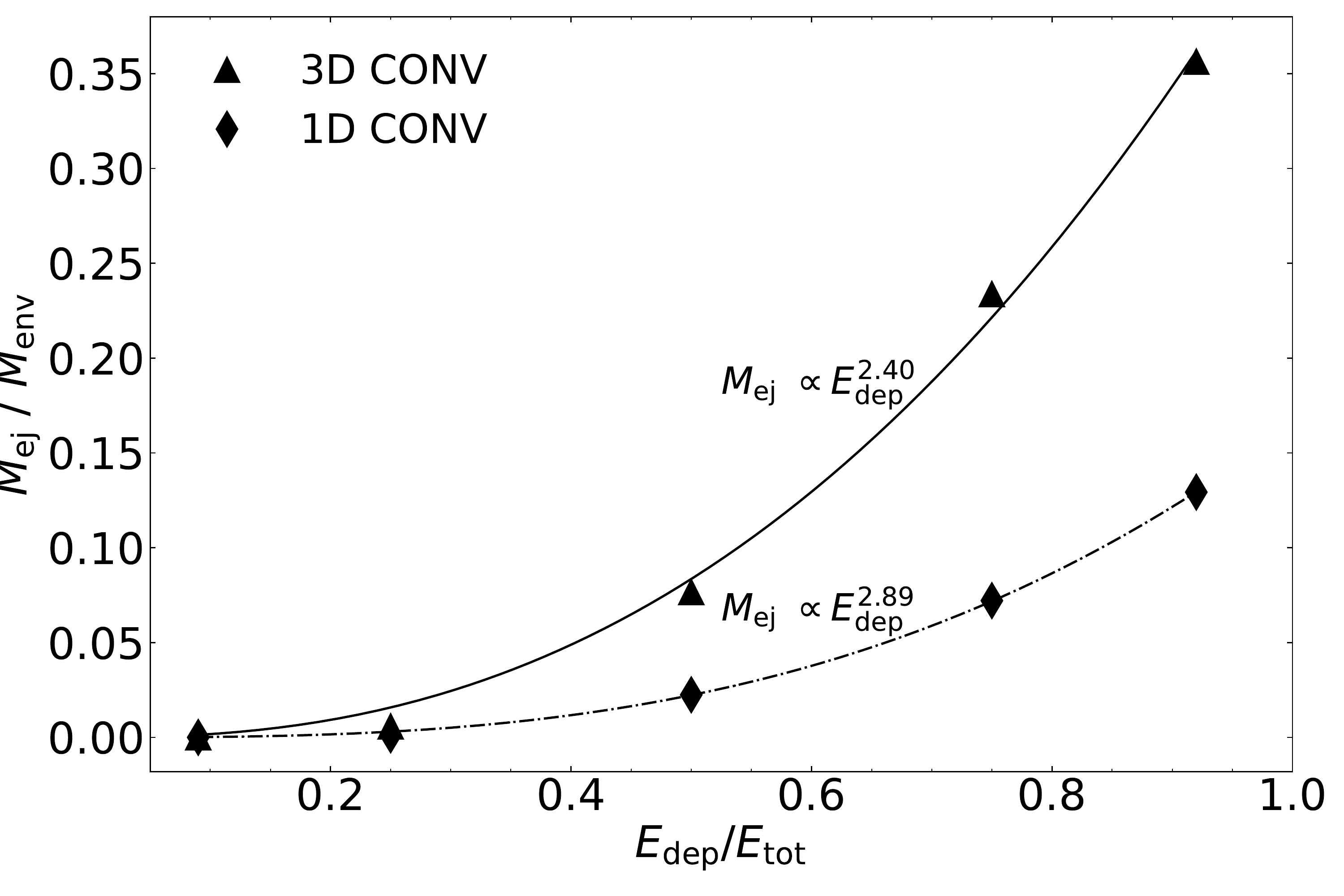}
\caption{Amounts of expected mass loss from the RSG envelope models as a function of injected energy, defined relative to the envelope's total energy.
Regardless of the deposited energy, 3D calculations including a convective envelope systematically result in more mass to be ejected than their 1D counterparts.
} \label{fig:Mej_Edep}
\end{figure}

In the model series CONV0.09--CONV0.92, we investigate the effects of varying the amount of energy deposited in both 1D and 3D.
The triangle symbols in Figure \ref{fig:Mej_Edep} depict the dependence of $M_{\rm ej}$ on $E_{\rm dep}$ in the 3D model series. In the parameter space explored in this work, where energy is injected over a duration shorter than the dynamical time, depositing more energy leads to more and more rapid mass loss. The solid black line shows a $M_{\rm ej} \propto E_{\rm dep}^{\gamma}$ power-law fit, with $\gamma_{\rm 3D} = 2.40$.
The results from the corresponding 1D runs are shown with diamond symbols, with $\gamma_{\rm 1D} = 2.89$. The power-law index from our 1D suite is very similar to $\gamma_{\rm poly} = 2.98$ found by \citet{Linial21} using $n = 3/2$ polytrope models.
However, without following the convective structures, the amounts of mass loss reported from the 1D simulations are 2--4 times lower than in 3D and the power-law index is substantially different.

\begin{figure}
\centering
\includegraphics[width=\columnwidth]{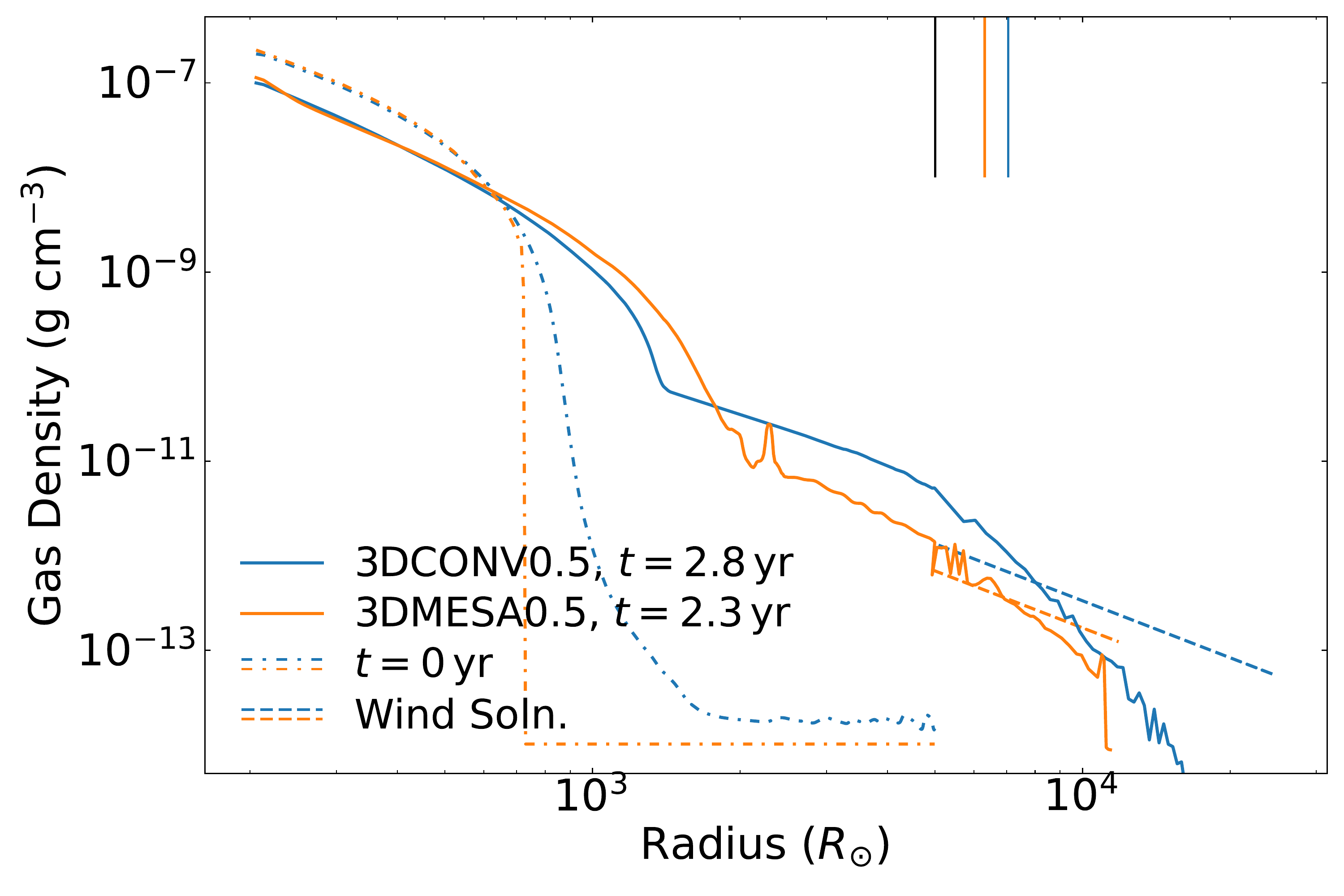}
\caption{Density profiles of the stellar envelopes and the ejecta from the 3DCONV0.5 and 3DMESA0.5 models (solid lines). The profiles are taken at the end of their mass loss episodes when the mass loss rates fall to zero, at 2.8 and 2.3\,years after energy deposition, respectively (see Figure \ref{fig:mass_loss_rate}). Dash lines denote the steady wind solution assuming a constant wind velocity of $v_{\rm w} = 65$\,km\,s$^{-1}$. The black vertical line marks the outer boundary of the \flash\ simulations. The colored vertical lines denote the half-mass radii of the escaped mass measured from the outer boundary. Density profiles immediately before energy deposition are plotted in dot-dashed lines as reference.
} \label{fig:ejecta_dens_profs}
\end{figure}

In the 1/3DCONV0.09 models, the majority of the injected energy is simply converted into gravitational binding energy, with minimal changes in the internal and kinetic energy content and negligible mass loss. In 1DCONV0.09, there is strictly zero mass loss. In 3D, $\sim$10$^{-4}$\,$M_{\odot}$ is ejected from the low-density top layer of the stellar envelope at $R \gtrsim 850$\,$R_{\odot}$. At this radius, radiation pressure starts to dominate and radiation hydrodynamical calculations will be required to determine the precise amount of mass loss.
Nevertheless, the non-zero mass loss in the 3D case indicates that the effects of convection are likely to modify predictions from 1D models \citep{Linial21}.

With the LR/HR models, we examine the effects of depositing the same amount of energy at different heating rates. In both 1D and 3D, we find that a larger deposition rate generally produces stronger shock acceleration and more ejected mass. 

\begin{figure*}
\centering
\includegraphics[width=\columnwidth]{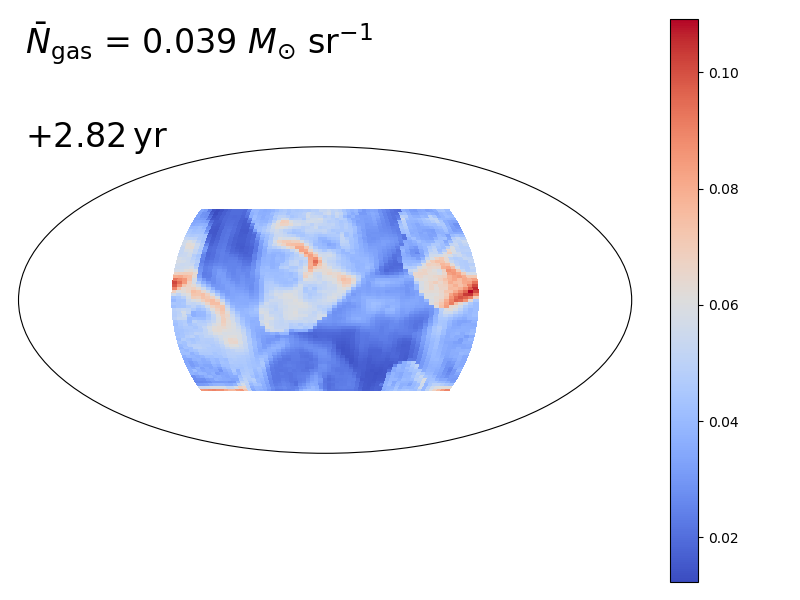}
\includegraphics[width=\columnwidth]{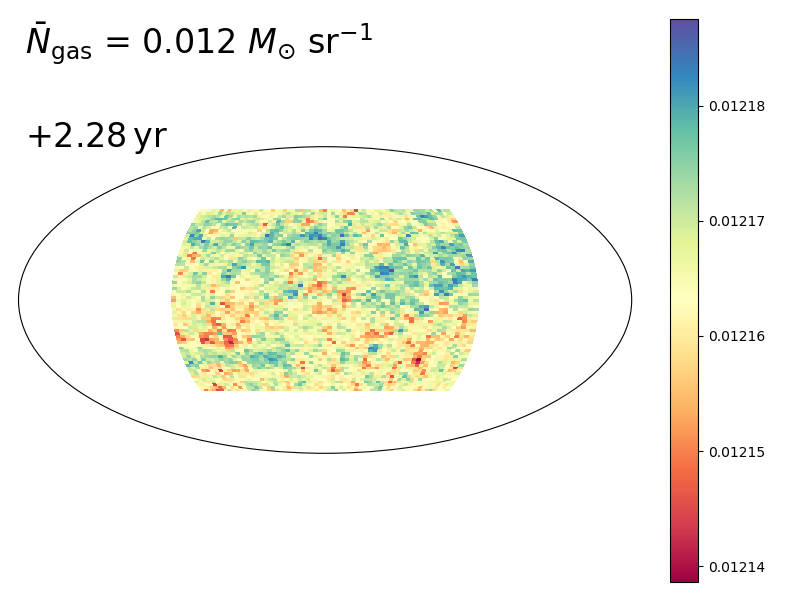}
\caption{Mollweide projection of the escaped mass from 3DCONV0.5 (left) and 3DMESA0.5 (right), expressed in $M_{\odot}$ per solid angle. These projections are taken at the end of the mass loss episode in each model, at the same moments as in Figure \ref{fig:ejecta_dens_profs}. Non-uniform shock propagation in the convective model leads to 10$\times$ variations in the column density at different lines of sight. In the initially static 3DMESA0.5 model, the column density variations are within $\lesssim$1\%. 
} \label{fig:ejecta_sky_dens}
\end{figure*}

\break

\subsection{Ejecta's Spatial Structures}
\label{sec:3dejecta}
In this section we quantify and visualize the structure of the ejecta in our models. In particular, we will analyze the density distributions of the ejected mass, examine the energetics to constrain the potential of the ejecta powering pre-SN outburst emission, and discuss how the ejecta could impact subsequent supernova emission.

The angle-averaged density profiles of the 3DCONV0.5 and 3DMESA0.5 models taken at the end of their mass loss episodes are plotted in Figure \ref{fig:ejecta_dens_profs}. 
The solid lines represent the stellar envelopes that remain in the simulation domain ($\leq R_{\rm outer}$) as well as the estimated density profiles of the ejecta ($> R_{\rm outer}$). The ejecta density profiles are computed by assuming that when gas clumps reach the outer simulation boundary, they continue to move outward at their instantaneous radial velocities, i.e., we ignore deceleration due to gravity once the gas clumps escape the simulation boundary.
The dashed lines show the $\rho \propto \dot{M} r^{-2} v_{\rm wind}^{-1}$ steady-state wind solutions using $\langle \dot{M} \rangle$ from Table \ref{tab:hydro_results} and with $v_{\rm wind}$ set to the time-averaged ejecta velocity of 65\,km\,s$^{-1}$.
The density profiles immediately before energy deposition are also shown for reference (dot-dashed lines).

Within 2--3 years after energy deposition, stellar materials can be launched up to $\approx$1.4$\times 10^{4}$\,$R_{\odot} = $10$^{15}$\,cm.
At $R \approx 2800-5000$\,$R_{\odot}$, \hlm{the model envelopes establish a wind-like structure with a $\propto r^{-2}$ density dependence.}
This is not surprising given that the mass loss rates are nearly constant when most of the mass loss is accumulated (see the plateaus in Figure \ref{fig:mass_loss_rate}).
\hlm{This wind-like layer differs from an actual wind in that it is still gravitationally bound by the progenitor.}
Above $\approx$5700\,$R_{\odot}$, the density profiles of the outer ejecta are steeper than the classical wind solution in both models.
In 3DCONV0.5, non-uniform gas acceleration in the convective stellar envelope gives rise to a slightly more extended ejecta structure and a more gradual decline. On the other hand, the density profile of the 3DMESA0.5 model has more of a shell-like appearance and a sharp drop-off.
Within $5000\,R_{\odot} \le R \le 1.2\times10^{4}\,R_{\odot}$, power-law fits to the density profiles yield $\propto r^{-5.3}$ and $\propto r^{-4.3}$ dependence for the 3DCONV0.5 and 3DMESA0.5 model, respectively.
If gas deceleration is taken into account, the ejecta structure would be even more compact.
\hlm{When a supernova goes off, its radiation signatures will be determined by the interactions of the supernova ejecta with both of the inflated envelope and the ejected outer layers.}

The KE$_{\rm ej}$ column in Table \ref{tab:hydro_results} summarizes the total kinetic energy contained in the ejecta material. Among the 3DCONV0.25--0.92 models, the ejecta carry about $10^{46-47}$\,erg of kinetic energy, consistent with the lower end of radiation energy observed in Type IIn SN precursors \citet{Strotjohann21}. 
Figure \ref{fig:mass_loss} shows that the majority of the mass loss is accumulated between 1 -- 2.5\,years following energy deposition regardless of the detailed model parameters. During this time, the radial velocities of the ejecta materials leaving the outer radius of the simulation domain fall from $\approx$100\,km\,s$^{-1}$ to $\approx$40\,km\,s$^{-1}$ monotonically. The decrease in velocity suggests that collisions of ejecta materials driven from the \emph{same} episode of energy deposition is unlikely.
The 20\,--\,100\,km\,s$^{-1}$ range of ejecta velocity observed in our simulations agrees with observed wind velocities from RSGs \citep{Goldman17}. 

The half-mass ejecta thickness of the 3DCONV0.5 (3DMESA0.5) model, measured from $R_{\rm out}$ outward, is $\Delta R_{\rm CONV, hm} = 2050$\,$R_{\odot}$ ($\Delta R_{\rm MESA, hm} = 1310$\,$R_{\odot}$).
We mark the radial locations of the half-mass radii with vertical lines in Figure \ref{fig:ejecta_dens_profs}.
If the ejecta's kinetic energy is thermalized from colliding with a \emph{previous}, similar episode of mass loss, the predicted luminosity would be $L_{\rm therm} = \epsilon\,{\rm KE}_{\rm ej} / (\Delta R_{\rm CONV, hm} / v_{\rm wind})$ $\approx \epsilon\,2 \times 10^{5}$\,$L_{\odot}$, where $\epsilon$ is the thermalization efficiency.
This estimate is consistent with the low-luminosity IIn SN precursors observed \citep{Strotjohann21,Jacobson-Galan22} \hlm{as well as the predicted luminosity of consecutive RSG mass eruptions estimated by \citet{Kuriyama21}}.

The 3D simulations allow us to examine the spatial distributions of ejecta materials. In Figure \ref{fig:ejecta_sky_dens} we show the column density map of the ejecta mass in the projected sky coordinate. In 3DCONV0.5, we see $\approx$10$\times$ variations in gas column density at different lines of sight.
Similar order-of-magnitude level variations in volumetric mass density are also observed across lines of sight.
These large fluctuations are produced from the non-uniform gas acceleration seeded by the initial convective motions in the envelope. In the initially static 3DMESA0.5 run where the envelope expansion is much more coherent, the line of sight variations in density are significantly lower at $\lesssim$1\%. 
Assuming a constant electron scattering opacity of $\kappa_{\rm es} = 0.33$, the integrated optical depths of the \hlm{angle-averaged} ejecta density profiles are $\tau_{\rm CONV} \approx 105$ and $\tau_{\rm MESA} \approx 43$.
Along the line of sight of the highest (lowest) column density, the integrated optical depth is $\tau_{\rm CONV, max} \approx 330.0$ ($\tau_{\rm CONV, min} = 31.0$).

In fact, models involving dense clumps embedded in steady winds have been invoked to explain the observed emission features of Type IIn SNe \citep{Chugai93,CD94,Smith09,Chugai18, Bevan20}. Based on simple analytical arguments, these dense clumps are typically also estimated to be $\sim$10$\times$ denser than the inter-clump medium. However, we emphasize that these models all assume a persistent, steady-state spherical wind superimposed with spherical dense clumps, unlike the highly asymmetric, filamentary, and episodic mass loss in our simulations.

Although it is beyond the scope of the current study, we note that the large variations in column density across lines of sight can have important implications in the shock dynamics, thermalization efficiency, and emission properties if the ejected materials is overrun by a subsequent episode of mass loss or a SN ejecta.

\section{Summary and Discussion}
\label{sec:summary}

Motivated by recent observations and the general lack of multi-dimensional models of pre-SN outbursts, we have presented a suite of 3D hydrodynamical simulations following the evolution of a convective RSG envelope upon energy injection. Our simulation suite was performed using 1D and 3D spherical geometry in \flash. Gas thermodynamics was followed using the \texttt{Helmholtz} EoS without radiation transport. We were agnostic toward the exact mechanisms that provide the energy of the outbursts. Energy was injected in the form of gas internal energy near the base of the model envelope. We explored energy deposition budgets that were lower than the gravitational binding energy of the envelope.
We focused on the limit where the injection timescale is shorter than the dynamical time of the envelope and the formation of shocks was expected.

We initialized our stellar envelope based on a \mesa\ RSG progenitor model. The \mesa\ model profiles were generated from a ZAMS mass of 15\,$M_{\odot}$, representative of the models used in many previous studies \citep{OM19,Leung20,Morozova20,Ko22}. Models in our simulation suite differ by the initial envelope structure (convective or static), amounts and rates of energy deposition, as well as geometry (1D and 3D). We explored energy deposition from 0.09 -- 0.92 times of the total energy of the envelope.
In our fiducial model with 0.5\,$E_{\rm tot}$ deposited, shock acceleration of the envelope led to an ejected mass of 0.5\,$M_{\odot}$, a time-averaged mass loss rate of 0.2\,$M_{\odot}$\,yr$^{-1}$, and a total kinetic energy of $1.7\times 10^{46}$\,erg in the ejecta.
In the following, we summarize the key results from our study and discuss their implications.

Convective motions intrinsic in the 3D model envelope can promote mass loss by allowing non-uniform acceleration of gas. They create low-density, fast-moving channels through which gas deep in the envelope can be launched to large radii (Figure \ref{fig:dens_slices_tile_2}, Figure \ref{fig:mass_loss_histogram}).
\citet{Leung20} also observed non-radial structures in similar 2D hydrodynamical models (see their Figure 2). However, since convection was not included in their initial model, they considered that their models may have underestimated the outward transport of energy if the convective flows were effective in helping to channel the deposited energy away.
The effectiveness of convective velocities in modifying the energy transport was indeed observed in our simulations.
The amount of ejected mass and mass loss rate are 2--3 times lower when energy is deposited onto an initially static model envelope regardless of dimensions. Since initializing hydrodynamical models directly from 1D stellar models before energy deposition is a common practice, it implies that some previous 1D studies may have underestimated the amount of potential mass loss.

Our results highlight that the 3D convective nature of massive star envelopes can be an important condition to consider in the understanding of pre-SN outbursts.
Following the wave heating hydrodynamics in \mesa,
\citet{WF22} recently found that the dissipation of gravity waves driven by nuclear burning is unlikely to cause significant mass loss. 
\hlm{In their setup, most of the wave energy is deposited at much smaller radii than our energy injection radius, at $\sim$10\,$R_{\odot}$. Unlike our simulations, their energy deposition timescale is several orders of magnitude longer than the local dynamical timescale at the much deeper injection radius.
Due to our relatively much larger $R_{\rm inner}$, we cannot examine energy deposition at smaller radii. 
Since the sound speed closer to the core is considerably higher, we expect that extending our simulations to smaller inner radii to capture the difference is feasible, but more computationally costly.
Since convective flows cannot be captured in 1D \texttt{MESA} models, it will be instructive to revisit their setup in 3D and verify whether the presence of the initial convective velocities in RSG envelopes impacts the outcome.
Our studies} demonstrate that besides the amount and duration, the location of energy deposition is another key parameter behind the launching of pre-SN outbursts.

The mass loss rates from various models vary by a factor of few during their energy deposition-induced mass loss episodes (Figure \ref{fig:mass_loss_rate}). The extent of variation is consistent with the time-varying mass loss rates inferred from multi-wavelength observations of some Type IIn SNe \citep{Fox13}.
Our range of mass loss rate $10^{-2} - 10^{-1}$\,$M_{\odot}$\,yr$^{-1}$ also overlaps with the estimated pre-explosion mass loss rates from an unbiased sample of IIn SNe, supporting the notion that typical IIn SNe may arise from energetic mass eruptions preceding the explosions  \citep{Kiewe12}.

The formation of shocks upon energy deposition and their traversal through the entire stellar envelope are in good general agreement with previous 1D hydrodynamical models \citep{Dessart10,Coughlin18,Fernandez18,Ko22,Linial21}. Since we opted to focus on non-terminal energy deposition with $E_{\rm dep} \lesssim E_{\rm bind}$, our model are qualitatively comparable to the partial ejection and inflated models in \citet{Dessart10}. 

In the range of energy scale we explored, depositing more energy results in higher mass loss rates, longer duration of mass loss, and a larger resultant ejected mass. The trend is in agreement with previous 1D hydrodynamical models \citep{Owocki19,Kuriyama20,Linial21}. Moreover, we found a scaling relation of $M_{\rm ej} \propto E_{\rm dep}^{2.89}$ in our 1D model series, very similar to the 1D numerical results of \citet{Linial21} with a $\propto E_{\rm dep}^{2.98}$ dependence. In 3D, we found a shallower scaling relation of $M_{\rm ej} \propto E_{\rm dep}^{2.40}$. We also point out that due to aspherical gas acceleration, one may not exactly reproduce the minimal mass-shredding explosion energy predicted in \citet{Linial21}.

With $E_{\rm dep} = 0.5\,E_{\rm bind}$, we found that a higher energy deposition rate led to more ejected mass. \citet{Ko22} have emphasized the necessary consideration of energy injection rate in the study of eruptive mass loss from massive stars. With the same amount of energy deposited at a lower rate, gravity effectively has a longer duration to act and decelerate the shocked gas, reducing the amount of gas above the escape velocity and the final mass loss. This phenomenon was verified in our 3DCONV0.5LR model. A similar trend was also observed in the \emph{high-energy models} in \citet{Leung20}.

RTIs quickly developed in the bottom of the initially static model envelope (Figure \ref{fig:dens_slices_tile_1}, top right). They smoothed out the density distribution and reduced the level of density inversion as compared to 1D. Later on, RTIs dissipated the density inversion completely (Figure \ref{fig:dens_prof_1D3D}, orange lines). Similar behaviors were already observed in previous 2D models \citep{Leung20}. The density structures of the envelope and ejecta have direct implications on the predicted light curves. It suggests that careful modeling in 3D may be required to provide more reliable observational constraints on progenitor/ejecta parameters.

Radial velocities of the ejecta decreased rather monotonically during the majority of the mass loss episode, from $\approx$100 to $\approx$40\,km\,s$^{-1}$. The velocity scale is consistent with previous studies. The decreasing trends of ejecta velocity imply that collisions between mass ejected from the same episode of energy release is unlikely. In other words, in order for the pre-SN outburst emission to be powered by collisions of ejected mass, recurrent outbursts may be required.
The thickness of the ejecta  is on the order of 1000--2000\,$R_{\odot}$ and the ejecta typically carry $10^{46-47}$\,erg of kinetic energy. If the ejecta's kinetic energy is thermalized into radiation over the thickness of the ejecta, its predicted luminosity is $\sim$$10^{4-5}$\,$L_{\odot}$, consistent with observations of IIn SN precursors \citep{Strotjohann21,Jacobson-Galan22}.

The density profiles of the ejecta are consistent with a wind-like mass shell (with a near-constant mass loss rate) and a sharp falloff. In other words, they can be described as a broken power law with $\propto r^{-2}$ in shell and $\propto r^{-4}-r^{-5}$ near the edge. 
Non-uniform gas acceleration in the 3D convective model leads to a 10$\times$ variations in the ejecta's column density and projected optical depth, comparable to the density contrast used in clumpy CSM models. The strong asymmetry and density variation have important implications in the dynamics, thermalization efficiency, and observed signatures of subsequent pre-SN and SN activities.

\section{Caveats and Future Directions}
\label{sec:caveats}
Our simulations were performed using the \texttt{Helmholtz} EoS. It assumes that the gas is fully ionized, and that radiation pressure is from an isotropic radiation field. In the top region of the envelope ($R \gtrsim 800$\,$R_{\odot}$), the gas temperature can fall below $10^{4}$\,K. Hydrogen recombination will lead to a state of partial ionization, and the abrupt changes in opacity and density will render the EoS highly inaccurate. Proper radiation hydrodynamics will be required to track the detailed radiation physics and thermodynamics in this top layer.
These potential improvements should affect a relatively small portion of mass since most of the additional mass loss channeled away by convective instabilities originates from deeper in the envelope. 
The general results on the effects of the convective envelope are likely robust.

In our study, we approached energy deposition on the RSG envelope in form of an agnostic, spherically-symmetric thermal heating. In reality, the outburst-driving mechanisms are triggered in the stellar envelope as an integral part of stellar evolution. The amount and the geometry of energy deposition are likely more complicated than our simple prescription. It will be meaningful to explore the responses of RSG envelopes to diverse modes of outburst-driving mechanisms in the future. In order to study the dynamics of recurrent outbursts, the outer radial boundary of the simulations will have to be extended to capture the fall-back and long-term behaviors of the mass ejecta. Higher spatial resolution will also be required at large radii to resolve the detailed dynamics of gas clumps subject to asymmetry and instabilities. 

Our hydrodynamical models were based on only one, albeit representative, \mesa\ stellar model as the initial conditions. Massive stars have a wide range of structures and binding energies. As shown in the model suite of \citet{Dessart10}, more massive ($M \gtrsim 25\,M_{\odot}$) stars tend to be more compact and the binding energy of the hydrogen envelope can be orders of magnitude higher.
Observationally, the diverse photometric and spectroscopic features observed among Type IIn SNe also point to a wide range of progenitor and environment properties \citep{Miller10,Kiewe12,Gangopadhyay20,Nyholm20,Ransome21}.
It will be informative to explore how mass loss and ejecta properties depend on the stellar models.

Since our models did not include radiation transport, we only estimated the radiative luminosity from potential ejecta interactions based on a simple order-of-magnitude energy argument. With radiation hydrodynamical calculations, one can constrain the outburst emission and directly compare their luminosities and variability timescales with pre-SN outburst observations. Our models provide the detailed 3D gas density structures surrounding the RSG progenitor. It will be valuable to model in follow-up studies the flash spectroscopy and supernova shock breakout signatures from the 3D ejecta structure.


\acknowledgments
\hlm{We are grateful to the anonymous referee for the constructive comments that improved the content of this paper.}
We thank Jared Goldberg for providing the \mesa\ red supergiant envelope model used in this study. We also thank Rodrigo Fern\'{a}ndez for sharing the source code used to extend the lower density limit of the \texttt{Helmholtz} equation of state in \flash.
This research project has benefited from interactions with Andrea Antoni, Jim Fuller, Shing-Chi Leung, Itai Linial, and William Schultz.
This research was funded by the Gordon and Betty Moore Foundation through Grant GBMF5076, and by NASA ATP-80NSSC18K0560 and ATP-80NSSC22K0725.
This research was supported in part by the National Science Foundation under Grant No. NSF PHY-1748958.
Use was made of computational facilities purchased with funds from the National Science Foundation (CNS-1725797) and administered by the Center for Scientific Computing (CSC). The CSC is supported by the California NanoSystems Institute and the Materials Research Science and Engineering Center (MRSEC; NSF DMR 1720256) at UC Santa Barbara.
\hlm{
The stellar profiles used as the initial conditions of the 3D convective equilibrium model is available on Zenodo under an open-source 
Creative Commons Attribution license: }
\dataset[doi:10.5281/zenodo.6857885]{https://doi.org/10.5281/zenodo.6857885}.

\software{
          \flash\ \citep{Fryxell00},
          Jupyter \citep{Kluyver16},
          Matplotlib \citep{Hunter07},
          \mesa\ \citep{Paxton2011,Paxton2013,Paxton2015,Paxton18,Paxton19},
          NumPy \citep{Oliphant06},
          SciPy \citep{Jones01},
          yt \citep{Turk11}.
          }

%

\vspace{5mm}

\bibliographystyle{aasjournal}
\bibliography{biblio}



\end{document}